\documentclass[useAMS,usenatbib,usegraphicx]{mn2e}

\def\msun{M$_\odot$}
\def\mstar{$M_*$}
\def\mgas{$M_{gas}$}
\def\fgas{$f_{gas}$}
\def\zsun{Z$_\odot$}
\def\mc{\multicolumn}
\def\cgs{erg~cm$^{-2}$sec$^{-1}$}
\def\mic{$\mu$m}
\def\arcsec{$^{\prime\prime}$}
\def\yeff{$y_{eff}$}
\def\ysun{$y_\odot$}

\def\ha{H$\alpha$}
\def\hb{H$\beta$}
\def\oii{[OII]$\lambda$3727}
\def\neiii{[NeIII]$\lambda$3869}
\def\oiiia{[OIII]$\lambda$4958}
\def\oiiib{[OIII]$\lambda$5007}
\def\oiii{[OIII]$\lambda$4958,5007}
\def\oiit{[OII]\small{3727}}
\def\neiiit{[NeIII]\small{3869}}
\def\oiiiat{[OIII]\small{4958}}
\def\oiiibt{[OIII]\small{5007}}
\def\hei{HeI\small{3889}}

\title[LSD: Mass, metallicity and gas at $z$$\sim$3.1]
{LSD: Lyman-break galaxies Stellar populations and Dynamics. 
I: Mass, metallicity and gas at $z$$\sim$3.1
   \thanks{Based on observations collected with ESO/VLT 
   (proposals 075.A-0300 and 076.A-0711), 
   with the Italian TNG, operated by FGG (INAF) at the 
   Spanish Observatorio del Roque de los Muchachos, 
   and with the Spitzer Space Telescope, operated by JPL (Caltech)
   under a contract with NASA.
   }
}

\author[F. Mannucci et al.]{
F. Mannucci$^1$\thanks{E-mail:filippo@arcetri.astro.it},
G. Cresci$^{1,2}$,
R. Maiolino$^{3}$,
A. Marconi$^{4}$,
G. Pastorini$^{4}$,
\newauthor
L. Pozzetti$^{5}$,
A. Gnerucci$^{4}$,
G. Risaliti$^{1,6}$, 
R. Schneider$^{1}$,
M. Lehnert$^{7}$, \&
M. Salvati$^{1}$\\
$^1$INAF - Osservatorio Astrofisico di Arcetri, 
   Largo E. Fermi 5, I-50125, Firenze, Italy\\
$^2$Max-Planck-Institut f\"ur extraterrestrische Physik (MPE), Giessenbachstr.1, D-85748 Garching, Germany\\
$^3$INAF - Osservatorio Astronomico di Roma, via di Frascati 33, I-00040 Monte Porzio Catone, Italy\\
$^4$Dip. di Astronomia e Scienza dello Spazio, Universit\`a di Firenze, 
   Largo E. Fermi 2, I-50125, Firenze, Italy\\
$^5$INAF - Osservatorio Astronomico di Bologna - Via Ranzani,1,
I-40127, Bologna, Italy\\
$^6$Harvard-Smithsonian Center for Astrophysics, 60 Garden Street, Cambridge, MA 02138, USA\\
$^7$GEPI, Observatoire de Paris, CNRS, University Paris Diderot, 5 Place Jules Janssen, F-92190 Meudon, France
}

\begin{document}

\date{Submitted 2009 February}

\pagerange{\pageref{firstpage}--\pageref{lastpage}} \pubyear{2007}

\maketitle

%-------------------------------------------------------------------------
\begin{abstract}
We present the first results of a project,
LSD, aimed at obtaining
spatially-resolved, near-infrared spectroscopy of a complete sample of
Lyman-Break Galaxies at $z$$\sim$3.
Deep observations with adaptive optics resulted in the detection of 
the main optical lines, such as \oii, \hb, and \oiiib,
which are used to study sizes, SFRs, morphologies,
gas-phase metallicities, gas fractions and effective yields.
Optical, near-IR and Spitzer/IRAC photometry is used to measure stellar mass.
We obtain that morphologies are usually complex, with the presence 
of several peaks of 
emissions and companions that are not detected in broad-band images.
Typical metallicities are 10--50\% solar, 
with a strong evolution of the mass-metallicity relation from lower 
redshifts.
Stellar masses, gas fraction, and evolutionary stages vary significantly 
among the galaxies, with less massive galaxies showing larger fractions of gas.
In contrast with observations in the local universe,
effective yields decrease with stellar mass and reach solar values 
at the low-mass end of the sample. 
This effect can be reproduced by gas infall with rates of the order of 
the SFRs.
Outflows are present but are not needed to explain the mass-metallicity 
relation.
We conclude that a large fraction of these galaxies
are actively creating stars after
major episodes of gas infall or merging. 
\end{abstract}

\begin{keywords}
Galaxies: abundances; Galaxies: formation; 
Galaxies: high-redshift;Galaxies: starburst
\end{keywords}
%
%=========================================================================
\section{Introduction}
\label{sec:intro}

Metallicity is one the most important properties of galaxies, and its study
is able to shed light on the details of galaxy formation.
It is an integrated property, not related to the present level
of star formation, but rather to the whole past history 
of the galaxy. In particular, metallicity is sensitive to the fraction of
baryonic mass already converted into stars, i.e., to the evolutionary stage
of the galaxy.
Also, metallicity is affected by the presence of infalls and outflows, i.e.,
by feedback processes and by the interplay between the forming galaxy
and the intergalactic medium.

The existence of a clear correlation between luminosity and
metallicity of galaxies is known from the late '70s \citep{Lequeux79}. 
It is now clear that local galaxies follow a well-defined mass-metallicity 
relation, where galaxies with larger stellar mass or larger circular velocity
have higher metallicities
\citep{Garnett02,Perez-Gonzalez03a,Pilyugin04,Tremonti04,Lee06,Panter08,Kewley08,Hayashi08,Michel-Dansac08,Liu08}.
The origin of the relation is uncertain because several effects can be, 
and probably are, active. 
It is well known that, in the local universe, starburst galaxies
eject a significant amount of metal-enriched gas into the intergalactic 
medium because of energetic feedback from exploding SNe (see, for example,
\citealt{Lehnert96a,Mori04} and \citealt{Scannapieco08}, 
and references therein).
Numerous studies have shown that outflows are already present at 
high redshifts \citep{Pettini01,Pettini02,Frye02,Weiner08}. 
Outflows are 
expected to be more important in low-mass galaxies, where the 
gravitational potential is lower and a smaller fraction of gas is retained.
As a consequence, higher mass galaxies are expected to be more metal rich
(see, for example, \citealt{Edmunds90} and \citealt{Garnett02}). 

A second possibile effect shaping the mass-metallicity relation
is related to the well know effect of ``downsizing'' 
(e.g., \citealt{Gavazzi96,Cowie96}), 
i.e., lower-mass
galaxies form their stars later and on longer time scales than more massive 
systems. At a given time, 
lower mass galaxies have formed a smaller fraction of their stars,
therefore are expected to show lower metallicities \citep{Ellison08a}.
Recently, \citet{Calura09} have explained the evolution of the 
mass-metallicity relation up to $z$=3.5 as due to an increase of the 
efficiency of star formation with galaxy mass, without invoking
differential galactic outflows.

Other possibilities exist, for example some properties of star formation,
as the initial mass function (IMF), could change systematically 
with galaxy mass \citep{Koppen07}.\\

All these effects have a deep impact on galaxy evolution, and the knowledge 
of their relative contributions is of crucial importance.
Different models have been built to reproduce the shape 
of the mass-metallicity relation in the local universe, and different 
assumptions produce divergent predictions
at high redshifts ($z>$2). 
 It is possible to test these predictions as metallicities can now be
measured at high redshifts \citep{Kobulnicky00,Pettini01,Pettini02,Maier06}.
To explore this issue several groups
have observed the mass-metallicity relation in the distant universe, 
around $z$=0.7 \citep{Savaglio05,Rodrigues08}, 
between 0 and 1.5 \citep{Cowie08}
between 0.5 and 1.2 \citep{Lamareille08,Perez08},
and at $z$$\sim$2 \citep{Erb06a,Halliday08,Hayashi08}. They have found 
a clear evolution with cosmic time, 
with metallicity decreasing with increasing redshift,
for a given stellar mass. 
\cite{Erb06a,Erb06b}
used near-IR long-slit spectra of a large number of LBGs at $z$$\sim$2
to estimate the average amount of extinction and estimate the intrinsic 
star formation rate (SFR).
By assuming that the Schmidt law holds at $z$=2 (see below), they also derive 
the fraction of baryonic mass in gas, and measure the effective yields.

There are several reasons to explore even higher redshifts.
The redshift range at $z$$\sim$3--4 is particularly interesting:  it is before 
the peak of the cosmic star formation density (see, for example, 
\citealt{Hopkins06,Mannucci07a}), only a small fraction 
($\sim$15\%, \citealt{Pozzetti07}) of the total stellar mass ($M_*$) 
has already been created, 
 is the redshift range where the most massive early-type galaxies 
are expected to form (see, for example, \citealt{Saracco03}),
the number of mergers among the galaxies is much larger
than at later times \citep{Conselice07,Stewart08}. 
As a consequence, it is above $z$$\sim$3
that predictions of different models tend to diverge significantly, 
and it is important to 
sample this redshift range observationally. The observations are really 
challenging because of the faintness of the targets and the precision 
required to obtain a reliable metallicity. Nevertheless, the new 
integral-field unit (IFU) instruments on 8-m class telescopes
are sensitive enough to allow for the project.

Metallicity at $z$$\sim$3 can be obtained by measuring the fluxes of the main
optical emission lines (\oii, \hb, \oiiib, \ha), whose 
ratios have been calibrated against metallicity
in the local universe (\citealt{Nagao06,Kewley08} and references therein).
Of course, this method 
can be applied only to line-emitting galaxies, i.e., 
to low-extinction, star-forming galaxies, whose lines can be observed 
even at high-redshifts (e.g., \citealt{Teplitz00a,Pettini01,Pettini02,Erb06a}). 
In contrast, the gas metallicity of more quiescent 
and/or dust extincted galaxies, like 
Extremely Red Galaxies (EROs, \citealt{Mannucci02}), 
Distant Red Galaxies (DRGs, \citealt{Franx03}), 
and Sub-mm Galaxies (SMGs, \citealt{Chapman05}), cannot be easily measured
at high redshifts. Stellar metallicities, revealed by absorption lines, 
can also be measured \citep{Gallazzi06,Panter08,Halliday08}, 
but at $z>$3 this is even more observationally challenging.\\

Together with metallicity, the dynamical properties of galaxies 
have a special role for their understanding. Dynamics is directly related to
the models of galaxy formation and it is the most direct way to probe the 
content of dark matter. 

Not much is known about dynamics of high redshift
galaxies as high-resolution, high-sensitivity, spectroscopic
observations are required.
\cite{Nesvadba06} and \cite{Stark08} observed highly 
lensed LBGs at $z$$\sim$3.1 to obtain the
velocity field. In these cases, the high spatial resolution on the source plane
is provided by the presence of the
gravitational lens, distorting the galaxy and stretching its apparent
dimensions to several arcsecs. \cite{Nesvadba06} found rotation 
velocities of 75 km/sec within the central lens-corrected 0.5 kpc, 
revealing the presence of a dynamical mass log(M/\msun)=9.3. 
\cite{Stark08} found a well ordered source, similar to local spirals.
\cite{Forster06a} obtained seeing-limited near-IR
integral field spectroscopy of a large sample galaxies 
at $z$$\sim$2. A significant fraction of 
this composite sample of spatially
extended, \ha-emitting galaxies shows dynamical properties consistent
with being the precursor of normal disk galaxies in the local universe.
Other recent works by \cite{Bouche07}, \cite{Genzel08} and \cite{Cresci09}
are addressing this problem, and the amount of 
accumulated information is rapidly growing.

\smallskip

Aimed at studying the high-redshift metallicity and dynamics, 
we started two projects,
AMAZE (Assessing the Mass-Abundance redshift Evolution)
and LSD (Lyman-break galaxies Stellar populations 
and Dynamics)\footnote{This acronym is not to be confused with two others
often used by astronomers, one of which is 
``Lens Structure and Dynamics'' by \citet{Koopmans03}.}.
For both projects we observed a sample of galaxies at $z$=3--4 with an 
IFU 3D spectrograph in order to derive their chemical and dynamical 
properties.
3D spectroscopy is a key technique for these studies as it allows to derive
the full velocity field of the galaxies, without the need to restrict the 
study to a slit. Also, integrated spectra of the galaxies can be 
obtained without slit losses.  
This is particularly important when the flux ratios between lines at
different wavelengths are derived, as differential losses can spoil the results.

For AMAZE, described in \cite{Maiolino08}, we selected 30 galaxies 
with deep Spitzer/IRAC photometry 
(3.6--8 $\mu m$), an important piece of information to derive a reliable
stellar mass. These galaxies were observed, in seeing-limited mode,
with the IFU spectrometer SINFONI on ESO/VLT.

The aim of LSD 
is not only to measure metallicities, but also to obtain 
spatially-resolved
spectra to measure dynamics and spectral gradients.
For this reason we used adaptive optics to improve the spatial resolution.
Also, we selected a complete, albeit small, sample of LBGs, in order to
obtain results that can be applied to the LBG population as a whole.
This is the first
paper about LSD, focusing on the chemical properties, while the dynamical 
properties will be examined in a forthcoming paper.

In the following, we adopt a $\Lambda$CDM cosmology with 
$H_0$=70 km/sec, $\Omega_m$=0.3 and $\Omega_\Lambda$=0.7.
At $z$=3.1, the universe is 2 Gyr old, 15\% of its current age.

%---------------------------------------------------------------
\begin{figure}
\centerline{\includegraphics[width=\columnwidth]{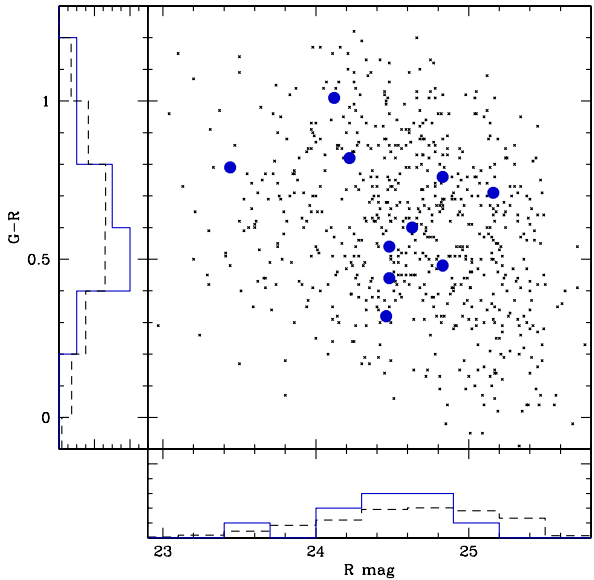} }
\caption{
Distribution in R magnitude and (G--R) color of the LSD targets
(solid blue dots)
compared to the spectroscopically confirmed targets of the Steidel et al.
(2003) sample with 2.5$<$z$<$3.3 (small dots). 
The solid blue histograms show the distributions of these 
quantities for the LSD sample, which are similar to those for the total 
parent sample (dashed histograms).
}
\label{fig:target}
\end{figure}

%-------------------------------------------------------------------
%-----------------------------------------------
\begin{table*}
\caption{Target list and summary of the observations} 
\label{tab:targets}
\begin{tabular}{lccccccccccc}
\hline
\hline
%Object    &\mc{2}{c}{R.A. (J2000) Dec.}&$R_{AB}$&z$_{em}$&\mc{2}{c}{Ref. Star}& Pix. scale &Exp. time \\
%            &             &             &      &        & Dist    &  R mag   &   (mas)    &  (min)    \\
   (1)    &      (2)      &     (3)     & (4) & (5)    &   (6)   &  (7) &   (8)         & (9) & (10) &   (11)  & (12)   \\
Object    &\mc{2}{c}{R.A. (J2000) Dec.}&$R_{AB}$&z$_{em}$&\mc{4}{c}{SINFONI}                  &   \mc{3}{c}{SPITZER}    \\
            &             &             &      &        & Dist   &R mag &  Scale        &Texp &T(1+3)&T(2+4)& N/A  \\
\hline
SSA22a-C30   &22:17:19.3 & +00:15:44.7 & 24.2 & 3.103  &35\arcsec& 13.0 & 50$\times$100 & 240 &  50   &   50   &   A  \\
SSA22a-C6    &22:17:40.9 & +00:11:26.0 & 23.4 & 3.097  &35\arcsec& 12.0 & 125$\times$250& 280 &  83   &  133   &   N  \\
SSA22a-M4    &22:17:40.9 & +00:11:27.9 & 24.8 & 3.098  &35\arcsec& 12.0 & 125$\times$250& 280 &  83   &  133   &   N  \\
SSA22b-C5    &22:17:47.1 & +00:04:25.7 & 25.2 & 3.112  &11\arcsec& 14.9 & 50$\times$100 & 240 &  83   &  133   &   N  \\
DSF2237b-D28 &22:39:20.2 & +11 55 11.3 & 24.5 & 2.933  &26\arcsec& 13.1 & 50$\times$100 & 240 &  96   &   96   &   A  \\
DSF2237b-MD19&22:39:21.1 & +11 48 27.7 & 24.5 & 2.611  &20\arcsec& 14.1 & 50$\times$100 & 200 &   -   &    -   &   -  \\
Q0302-C131   &03:04:35.0 &--00:11:18.3 & 24.5 & 3.235  &26\arcsec& 13.4 & 50$\times$100 & 240 & 332   &  133   &   N  \\
Q0302-M80    &03:04:45.7 &--00:13:40.6 & 24.1 & 3.414  &15\arcsec& 14.5 & 50$\times$100 & 240 & 166   &  399   &   N  \\
Q0302-C171   &03:04:44.3 &--00:08:23.2 & 24.6 & 3.328  &33\arcsec& 14.6 & 50$\times$100 & 240 & 332   &    -   &   N  \\
Q0302-MD287  &03:04:52.8 &--00:09:54.6 & 24.8 & 2.395  &13\arcsec& 15.0 & 50$\times$100 & 160 & 249   &    -   &   N  \\
\hline
\hline
\end{tabular}
%\flushleft
Columns: (1-3): Target name and coordinates; (4): Target R-band magnitude; (5): Redshift from emission lines;
(6-7): Guide star distance and R-band magnitude; 
(8-9): spatial scale (mas) and total exposure time (min) of the 
SINFONI observations;
(10-11): exposure time (min) for Spitzer/IRAC observations in channels 1+3 and 2+4; (12) New (N) or Archive (A) Spitzer/IRAC data.

\end{table*}

 % targets and observations

%=========================================================================
\section{Sample selection}
\label{sec:sample}

Our sample of LBGs was extracted from the \cite{Steidel03} catalog,
which contains about 1000 spectroscopically-confirmed LBGs, 
selected in 17 fields down
to a limiting magnitude of $R_{AB}=25.5$. 
The parent catalog was searched for the presence of bright stars (R$<$15)
within $\sim$35\arcsec\ of each source, resulting in the selection of 
$\sim$100 sources. 
The presence of a bright  foreground star is needed to drive the
adaptive optics module and obtain a resolution comparable with the 
diffraction limit of a 8m class telescope, about 0.1\arcsec\ in the near-IR.
The best combinations of star brightness and distance from the
target were selected to obtain the final sample of 10 objects presented in
Table~\ref{tab:targets}.

We emphasize that no other property was used to select the 
targets, thus the present sample, albeit small, should be representative of
the LBG population in the \cite{Steidel03} sample.
This is shown in figure~\ref{fig:target}, which compares the distributions
of R-band magnitude and (G--R) color of the selected targets
with those of the parent population of LBGs. The distributions of both 
quantities are similar, with no apparent biases. 

\smallskip

In order to check the possible presence of Active Galactic Nuclei,
we inspected the available X-ray observations of these galaxies.
For two objects, DSF2237b-D28 and DSF2237b-MD19, no X-ray observation
is available. The other sources lay in the field of a deep {\em Chandra}
observation (SSA22) and in that of a quasar observed with {\em XMM-Newton}.
No source is detected in the 0.5-10~keV range. In all cases, 
the off-center position of the sources, and/or the short
exposure time imply a relatively high upper limit for the X-ray flux, which,
given the high redshift of the sources, corresponds to X-ray luminosities
larger than $\sim10^{43}$~erg~s$^{-1}$. 
Although these limits cannot rule out
the presence of an AGN, there is no direct 
evidence for it. In all cases 
we can exclude a dominant, quasar-like AGN contribution.

%=========================================================================
\section{Observations and data reduction}
\label{sec:obs}

The selected objects have been observed with several instruments in order to
obtain a complete data set.

%-------------------------------------------------------------------------
\subsection{Integral field spectroscopy}
\label{sec:sinfoni}

The targets were observed at ESO/VLT by using the SINFONI instrument
\citep{Eisenhauer03,Bonnet04},
an IFU spectrograph fed by an adaptive optics module
with a Natural Guide Star (NGS).
The IFU splits the incoming light into several adjacent slits,
and a spectrum is obtained for each
location in the field-of-view. Several spatial scales are available, 
corresponding to different spatial resolutions. For 8 of the 10 targets we used 
the 50$\times$100 mas/pixel scale, providing a field-of-view 
of 3\arcsec$\times$3\arcsec.
Such a scale is a good match of the theoretical diffraction limit in the K-band 
on the VLT telescope, which is $\sim$70 mas, and the field-of-view is 
appropriate to observe galaxies of dimensions of $\sim$1\arcsec.
Two objects constitute a pair of interacting galaxies at a distance of about
2\arcsec, in this case we used a larger scale, 125$\times$250 mas/pixels 
with a total field-of-view of 8\arcsec$\times$8\arcsec.

In all cases we used the H+K grating, providing a simultaneous coverage
of these two near-IR bands (1.5--2.4\mic), with resolution
$\lambda/\Delta\lambda\sim$1400 in the H band and 
$\sim$1800  in the K band.

Observations were divided into independent sequences of about 45 min each, 
due to the
limitation of the ESO service observing strategy based on observing blocks (OBs)
of about 1 hour of total time.
During each OB, an ABBA nodding was used, putting the target in two positions
of the field-of-view about 1.5\arcsec\ apart. This is a very efficient way to
observe compact objects as no time is lost in observing the sky. Nevertheless,
in principle
this could introduce some self-subtraction of the wings of the targets
more extended than 1.5\arcsec.

A maximum seeing of 0.8\arcsec\ was requested to carry out the observations, 
and an image of the reference star was observed at the beginning of each OB
to monitor the conditions of the atmosphere during the observations. 

Data were reduced by using the ESO pipeline \citep{Modigliani07} with the
improved sky subtraction described by \cite{Davies07a}. 
After flat-fielding, sky-subtraction, correction for distortions and
wavelength calibration, the 2D data of each OB 
are mapped 
into a 3D data cube with dimensions $32\times64\times2048$ pixels. 
The cubes of the different OBs are then combined together by measuring the
relative offsets from the detected centroid of emission. 
We note that using the coordinates
in the image header often does not provide the requested 
precision ($<$0.1\arcsec),
as the measured pointing uncertainties are of the order of 0.2\arcsec.

%-------------------------------------------------------------------------
\subsection{Photometric and morphological data}
\label{sec:photom}

The original dataset of optical photometry in the UGR bands
by \cite{Steidel03} is available for all the objects.
Such a limited wavelength range does not allow to obtain reliable
stellar masses by fitting the spectral energy distribution (SED). 
The optical data must be complemented by near-IR photometry.
J and K-band photometry of 4 objects was published by 
\cite{Shapley01}. We have observed the remaining 6 objects
in the J and Ks filters
with the NICS camera \citep{Baffa01} on the Italian 3.6m telescope TNG.
Exposure times of 1--1.5 hours in J and 2--3.5 hours in Ks were used,
according to the expected near-IR magnitudes. NICS data were reduced by
the automatic pipeline 
SNAP\footnote{http://www.arcetri.astro.it/$\sim$filippo/snap}.\\

Photometry at longer wavelengths (3.6--8\mic) 
is available in the Spitzer/IRAC archive
\citep{Werner04,Fazio04}
for some of the objects in the SSA22 field. 
For the remaining objects we obtained targeted observation with
Spitzer/IRAC during cycle 7, 
as listed in Table~\ref{tab:targets}. 
We used the post-BSD products of the pipeline, and 
measured aperture photometry.\\

Finally, HST will observe these objects after SM4, both 
in the optical with ACS and in the
near-IR with WFC3, providing accurate photometry and broad-band
morphologies. In conclusion, these objects will have a complete set of 
photometric and spectroscopic observations with  detailed morphological information
at optical and near-IR bands, and will be suitable for detailed studies.

%=========================================================================
\section{Results and measured quantities}

%----------------------------------------------------------------------
\subsection{The detected lines}
\label{sec:lines}
%-----------------------------------------------
\begin{table}
\caption{Properties of the detected lines}
\label{tab:flux}
\begin{tabular}{lccccc}
\hline
\hline
  (1)      &  (2)    &   (3)   &    (4)          &   (5)      \\
Object     &line    & $z$      & Flux($^a$)      & Fact($^b$)\\
\hline
SSA22a-C30 &   \oiit & 3.1024 &  0.30 $\pm$ 0.12 & 3.18 \\ %OK
           &    \hb  & 3.1030 &  0.26 $\pm$ 0.10 &      \\ %OK
           & \oiiiat & 3.1034 &  0.34 $\pm$ 0.10 &      \\ %OK
           & \oiiibt & 3.1026 &  1.28 $\pm$ 0.22 &      \\ %OK
\hline
SSA22a-C6  &   \oiit & 3.0966 &  1.21 $\pm$ 0.38 & 1.09 \\ %OK
           &    \hb  & 3.0970 &  0.76 $\pm$ 0.45 &      \\ %OK
           & \oiiiat & 3.0966 &  2.26 $\pm$ 0.78 &      \\ %OK
           & \oiiibt & 3.0968 &  5.50 $\pm$ 0.62 &      \\ %OK
\hline
SSA22a-M4  &   \oiit & 3.0961 &  1.46 $\pm$ 0.23 & 1.06 \\ %OK
           &    \hb  & 3.0976 &  0.69 $\pm$ 0.25 &      \\ %OK
           & \oiiiat & 3.0965 &  1.08 $\pm$ 0.30 &      \\ %OK
           & \oiiibt & 3.0978 &  3.66 $\pm$ 0.49 &      \\ %OK
\hline
SSA22b-C5  & \neiiit & 3.1098 &  0.38 $\pm$ 0.14 & 1.57 \\ %OK
           &    \hb  & 3.1122 &  0.50 $\pm$ 0.14 &      \\ %OK
           & \oiiiat & 3.1120 &  1.01 $\pm$ 0.22 &      \\ %OK
           & \oiiibt & 3.1121 &  3.28 $\pm$ 0.24 &      \\ %OK
\hline
D2237b-D28 &  \oiit  & 2.9352 &  0.74 $\pm$ 0.16 & 1.84 \\ %OK
           & \oiiiat & 2.9328 &  0.50 $\pm$ 0.16 &      \\ %OK
           & \oiiibt & 2.9324 &  1.86 $\pm$ 0.36 &      \\ %OK
\hline
D2237b-MD19&\hb      & 2.6093 &  0.60 $\pm$ 0.16 & 1.95 \\ %OK
           &\ha      & 2.6108 &  2.96 $\pm$ 0.46 &      \\ %OK
\hline
Q0302-C131  &  \oiit & 3.2349 &  0.80 $\pm$ 0.32 & 1.29 \\ %OK
            &   \hb  & 3.2346 &  0.46 $\pm$ 0.16 &      \\ %OK
            &\oiiiat & 3.2341 &  0.90 $\pm$ 0.20 &      \\ %OK
            &\oiiibt & 3.2346 &  2.62 $\pm$ 0.52 &      \\ %OK
\hline
Q0302-M80   &  \oiit & 3.4136 &  0.56 $\pm$ 0.18 & 1.23 \\ %OK
            &   \hb  & 3.4138 &  0.40 $\pm$ 0.12 &      \\ %OK
            &\oiiiat & 3.3990 &  0.42 $\pm$ 0.20 &      \\ %OK
            &\oiiibt & 3.4137 &  1.32 $\pm$ 0.26 &      \\ %OK
\hline
Q0302-C171  &   \oii & 3.3287 &  0.56 $\pm$ 0.24 & 1.92 \\ %OK
            &    \hb & 3.3289 &  0.18 $\pm$ 0.08 &      \\ %OK
            & \oiiia & 3.3280 &  0.28 $\pm$ 0.08 &      \\ %OK
            & \oiiib & 3.3279 &  1.10 $\pm$ 0.12 &      \\ %OK
\hline
\hline
\end{tabular}
Columns. 
(4): line flux inside an aperture of 0.35\arcsec with correction for the
PSF, units of $10^{-17}$\cgs; 
(5): scaling factor to total flux 
\end{table}

The final SINFONI data cubes were searched for the position of the 
brightest lines,
either \oiiib\ or \ha, to define the spatial position of the target.
The main optical lines (\oii, \hb, \oiiib\ and \ha) are  detected in 9 out 
of 10 galaxies, the only exception being Q0302-MD287. 
The reasons for this non detection are not clear. It is possible that its
emission lines fall below our detection threshold, or that they 
coincide with bright sky lines. 
The SED fitting described in sect.~\ref{sec:mass}
points toward the existence of a large amunt of dust in this object, with $A_V$ between 0.9 and 1.1 mag.
As a consequence, it is also possible that the optical lines are not detected because of the presence of
large amount of dust extinction.

The choice of the aperture for the extraction of the spectra is a important step.
Measuring metallicity requires
a good knowledge of line ratios, which are best obtained inside small apertures
on the central regions of the galaxies, where the signal-to-noise is higher.
In contrast, measuring total SFRs requires larger apertures to maximize
the fraction of line emission recovered by aperture photometry.
%To obtain values of the line fluxes to measure all these quantities, 
%we followed the following strategy.
To measure all these quantities, we extracted complete spectra 
inside a circular aperture of 
0.35\arcsec\ of diameter, corresponding to 7 pixels, and an aperture correction
for a point source was applied to recover the emitted flux of the central
part of the galaxies. 
We chose a circular aperture, instead of more
complex apertures based on object morphology and surface brightness, in
order to avoid any problem with possible variations of the PSF with
wavelength and a non perfect spatial alignment of the spectra.
The fraction of total flux recovered by this procedure was estimated 
by computing
the curve of grow of the photometry of the brightest line as a function of 
aperture radius. A correction for the missing flux is applied when the ``total''
quantities are needed.
The results are shown in Table~\ref{tab:flux}, listing the line fluxes inside the
circular aperture and the correcting factor to total fluxes. 
Except for one galaxy, SSA22a-C30, 
composed by several clumps of emission, for all the other galaxies the
tabulated fluxes comprise more then half of the total flux.
\\

%----------------------------------------------------
\begin{figure*}
\includegraphics[width=3.8cm,bb=0 -80 814 447] {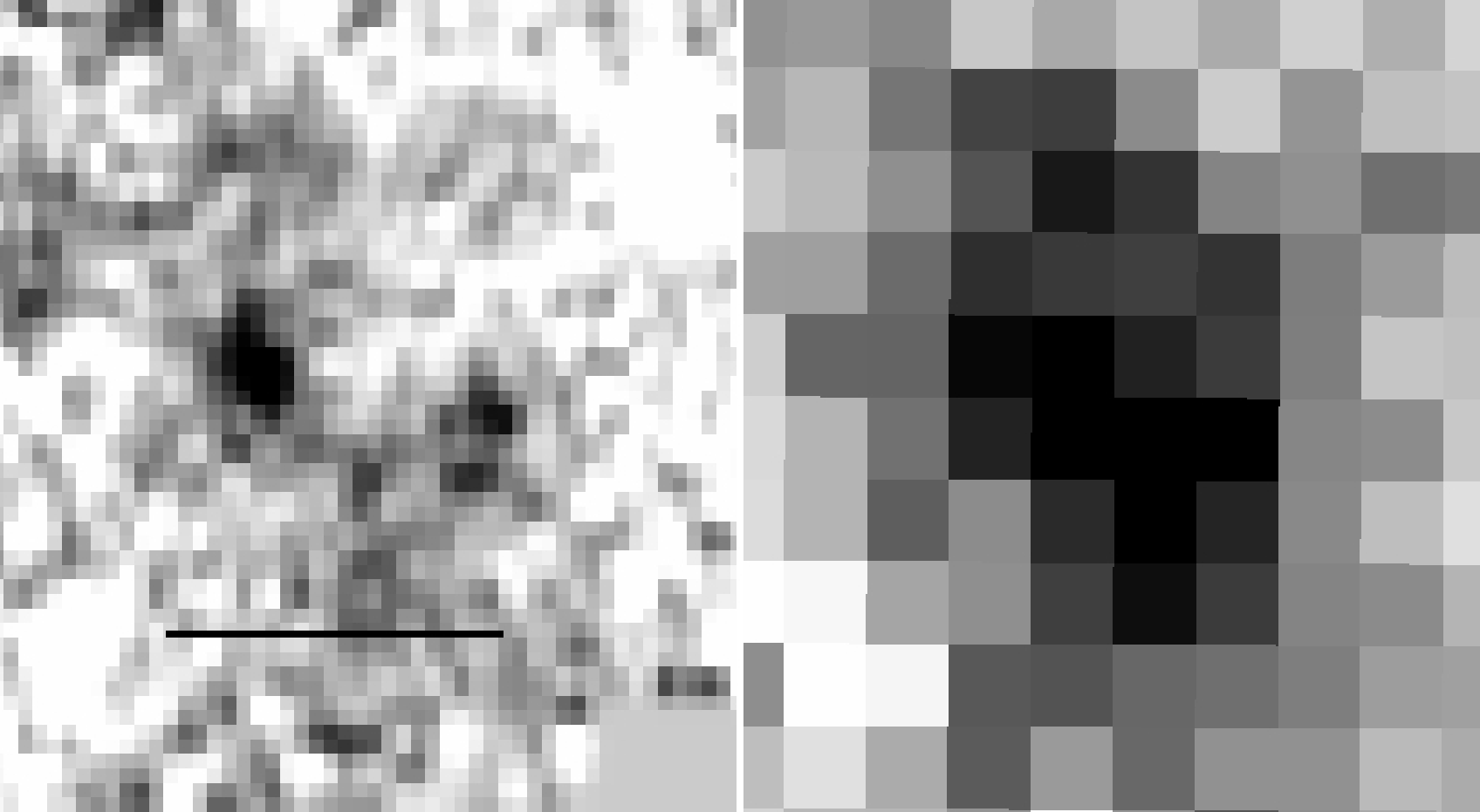} 
\includegraphics[width=4.4cm,height=3.5cm]     {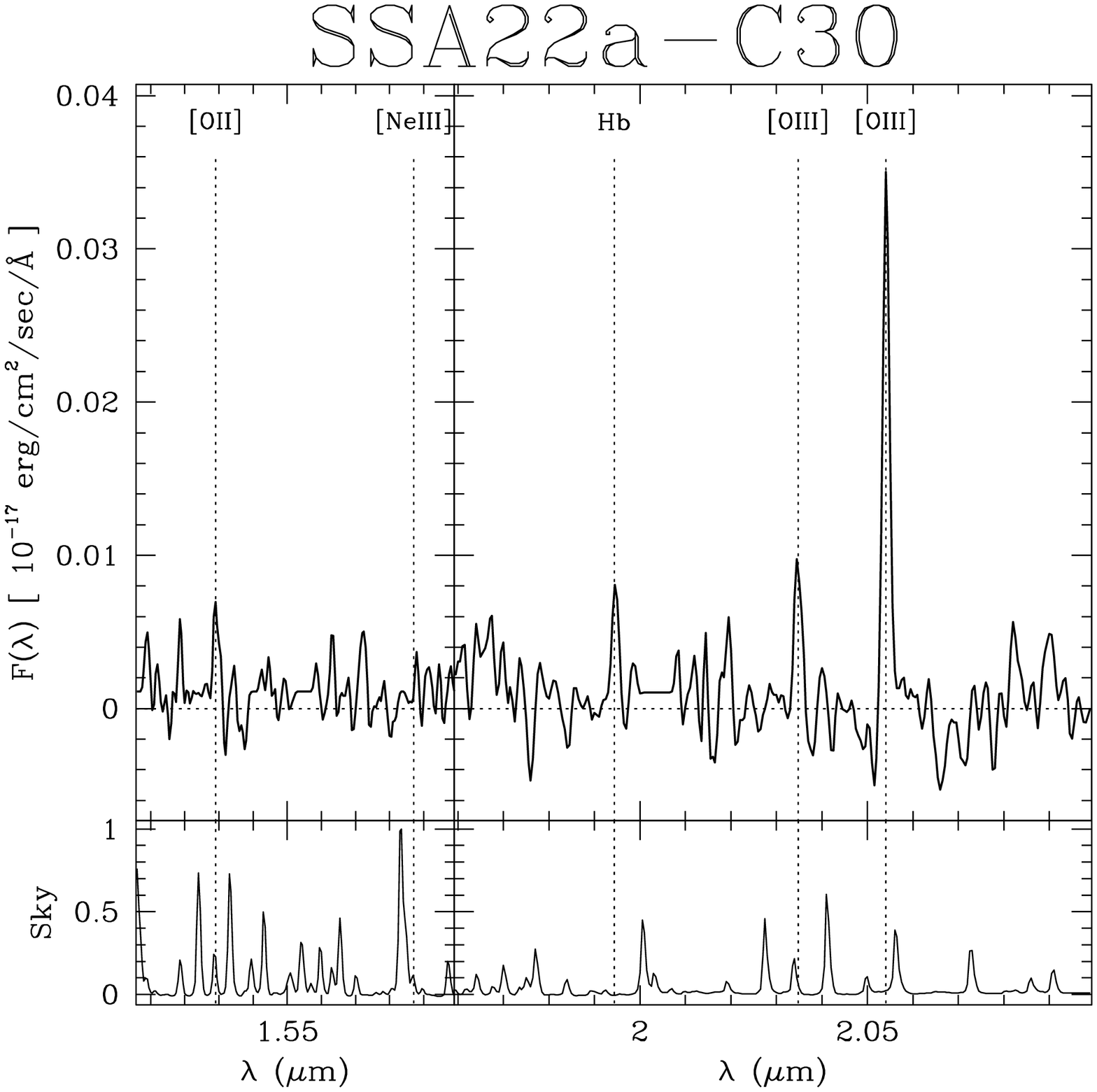} 
\hspace*{2mm}
\includegraphics[width=3.8cm,bb=0 -80 814 447]{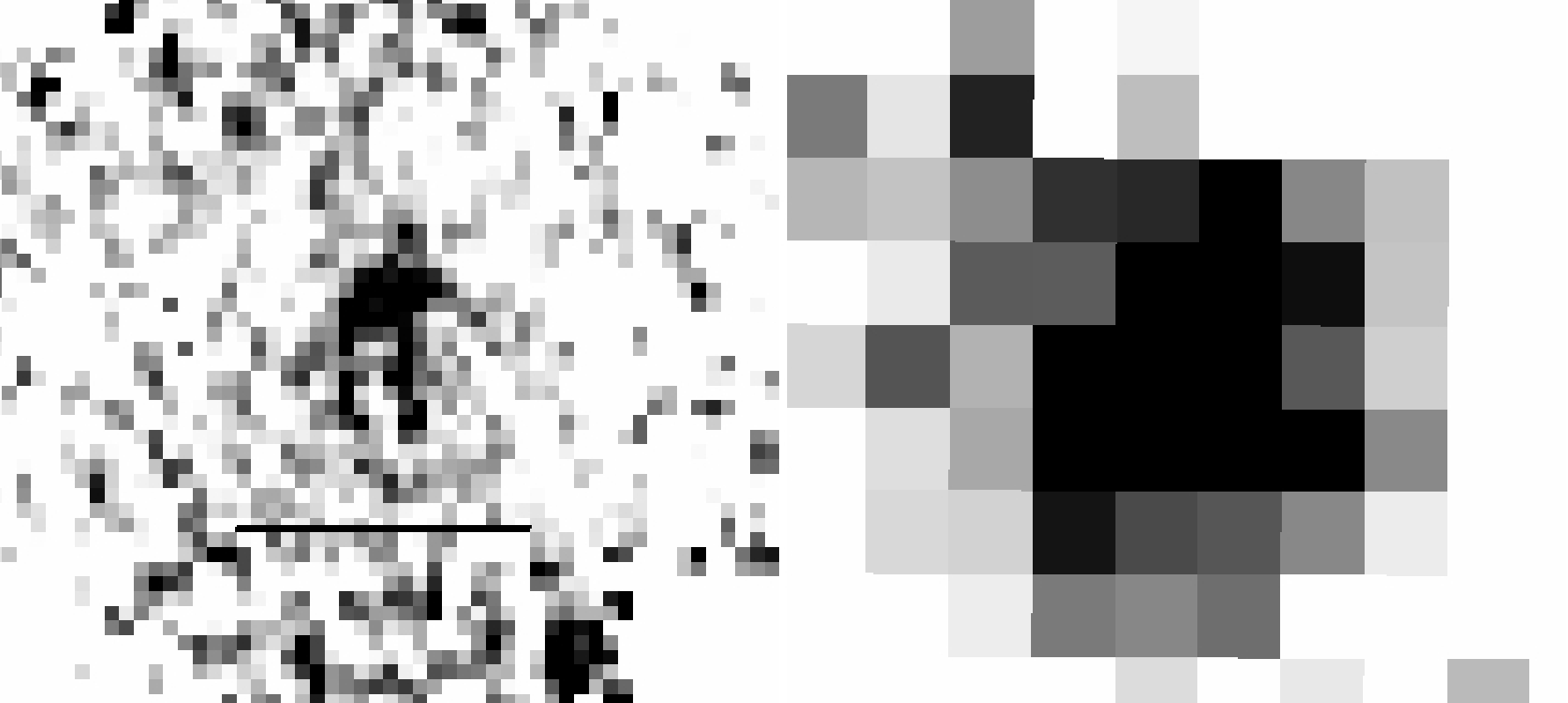} 
\includegraphics[width=4.4cm,height=3.5cm]    {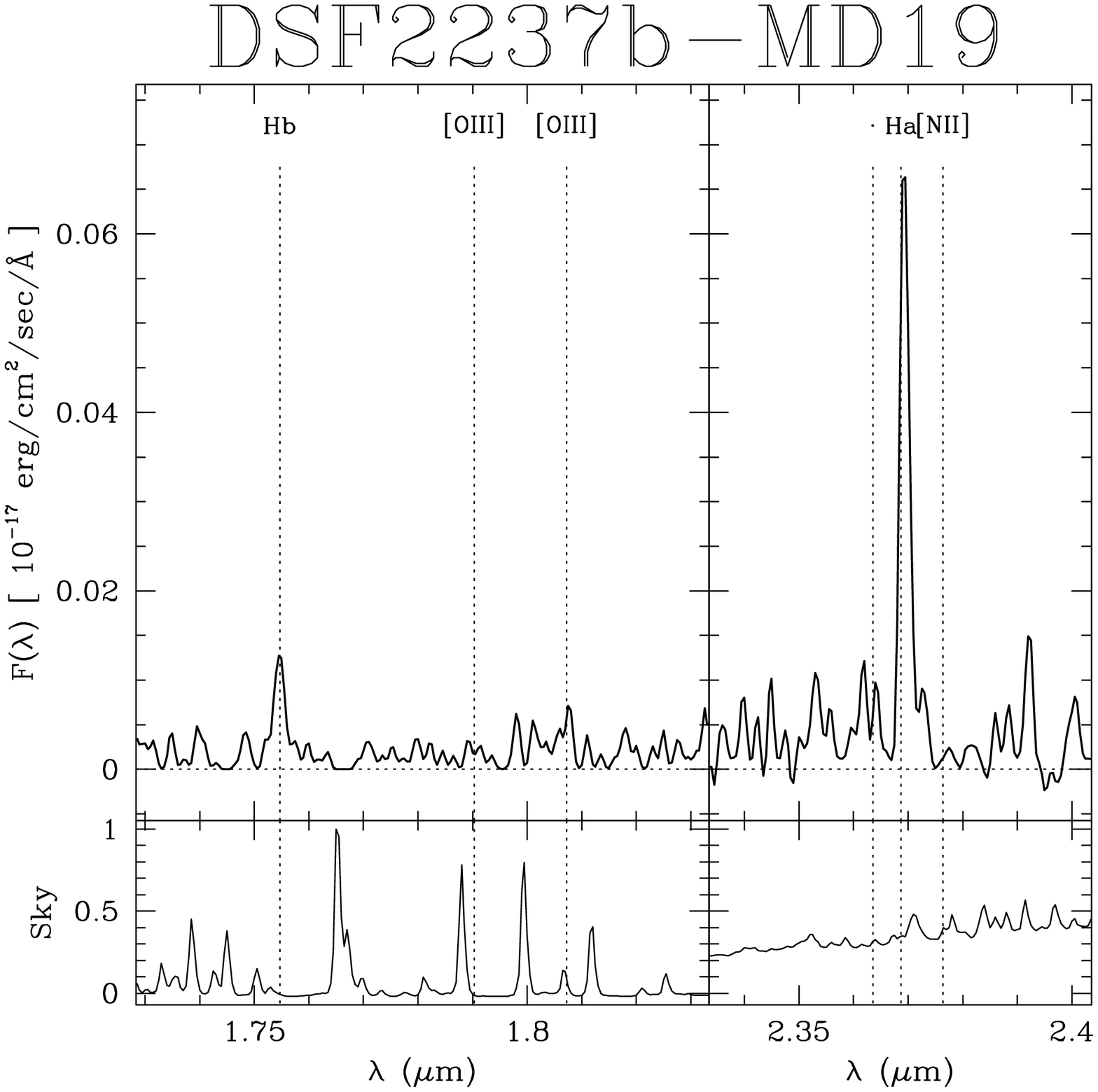} 
\\ \smallskip
\includegraphics[width=3.8cm,bb=0 -350 662 470]{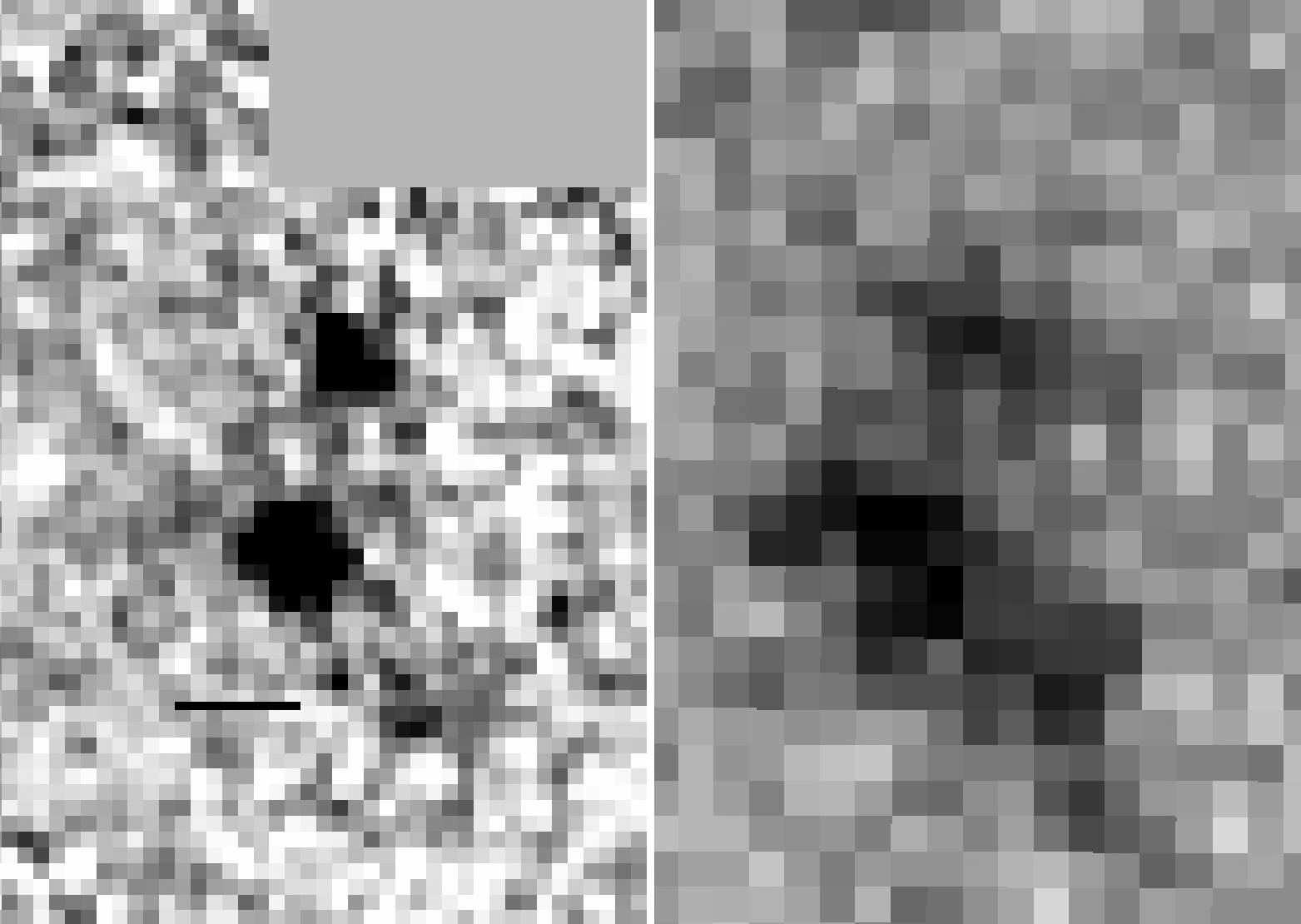} 
\includegraphics[width=4.45cm,height=7.4cm,bb=0  -450 592 718]{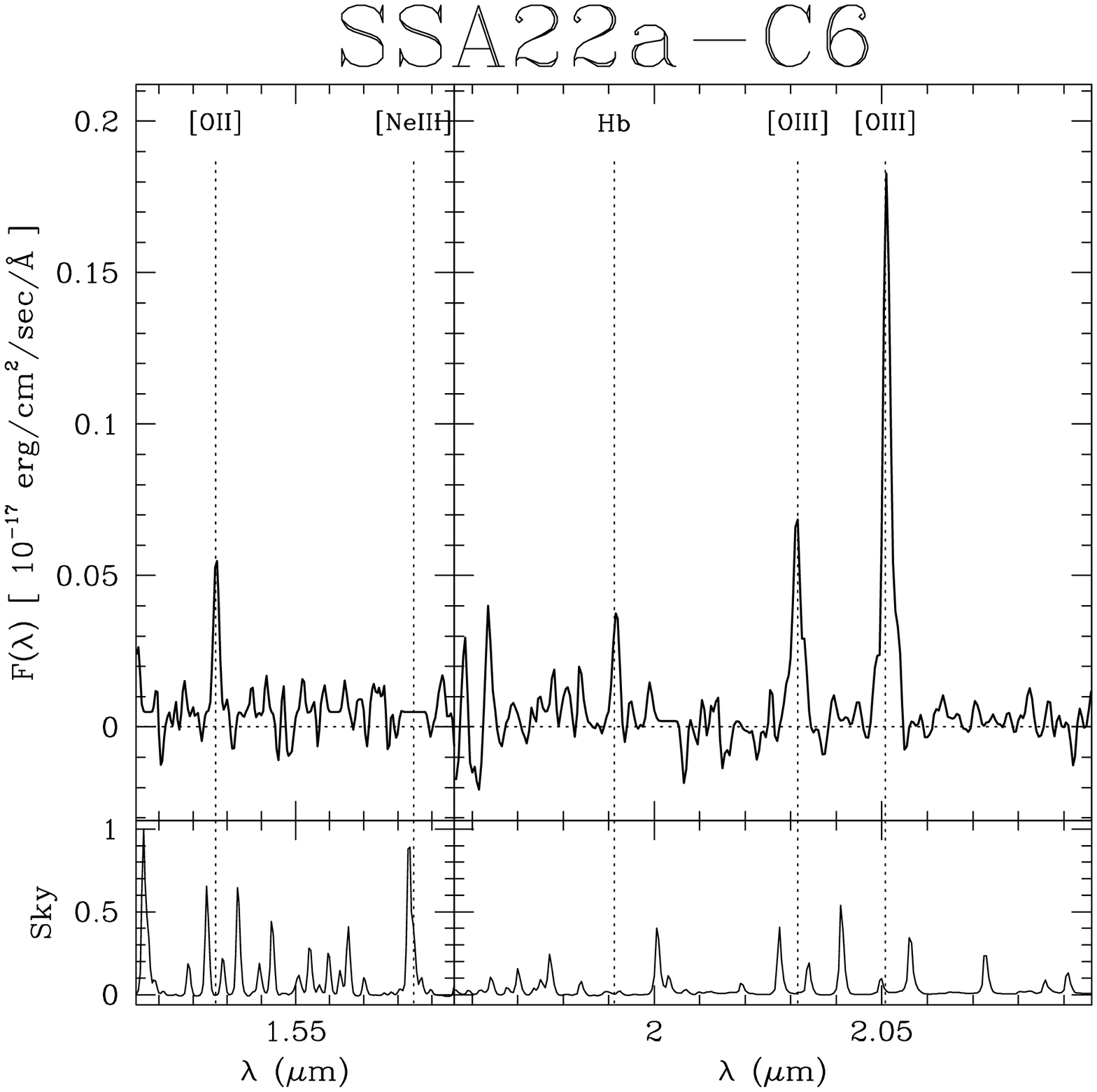} \hspace*{-45mm}
\includegraphics[width=4.4cm,height=7.4cm,bb=18 144 592 1268]{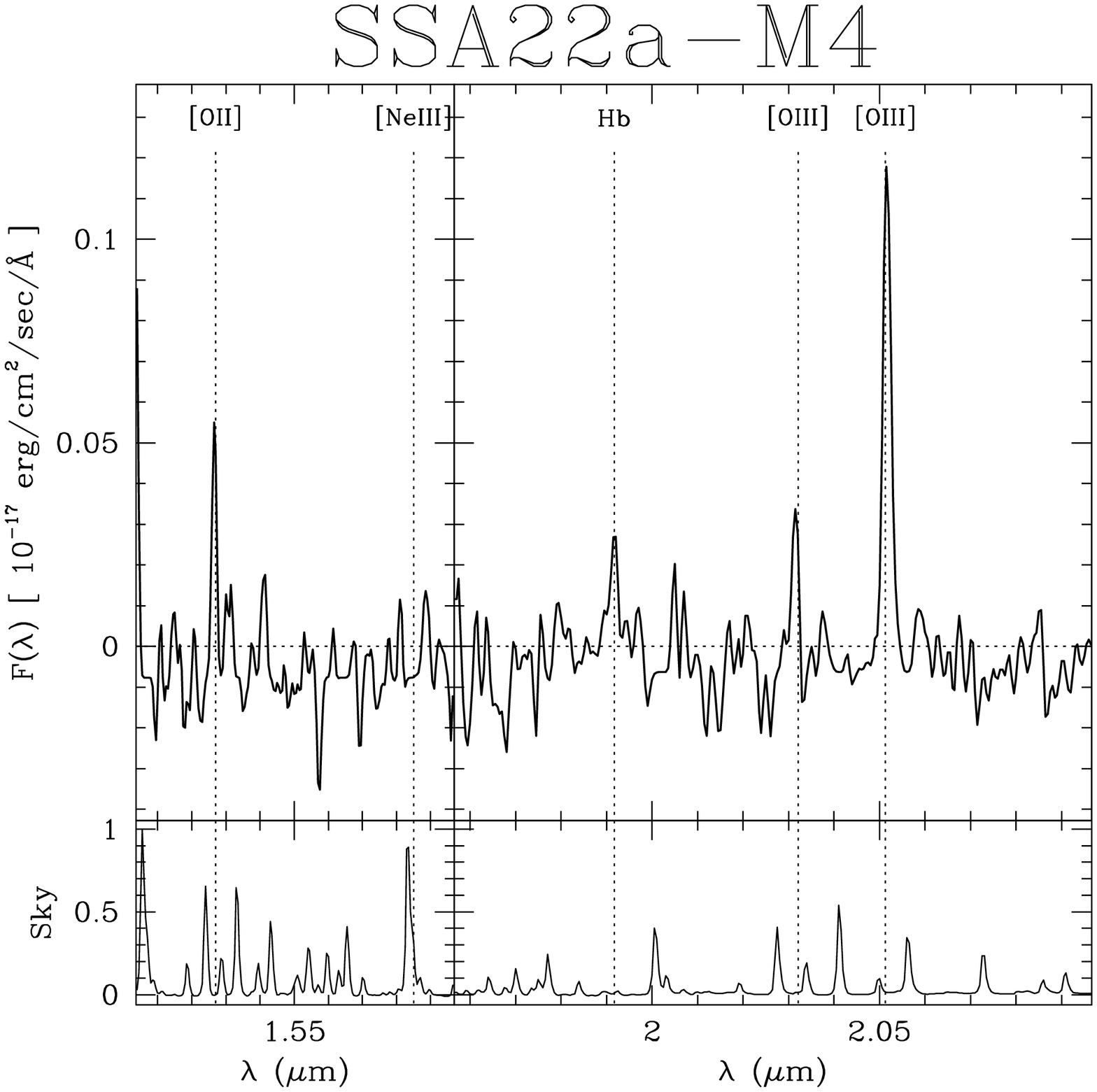} 
\hspace*{2mm}
\includegraphics[width=3.8cm,bb=0 -460 757 422]{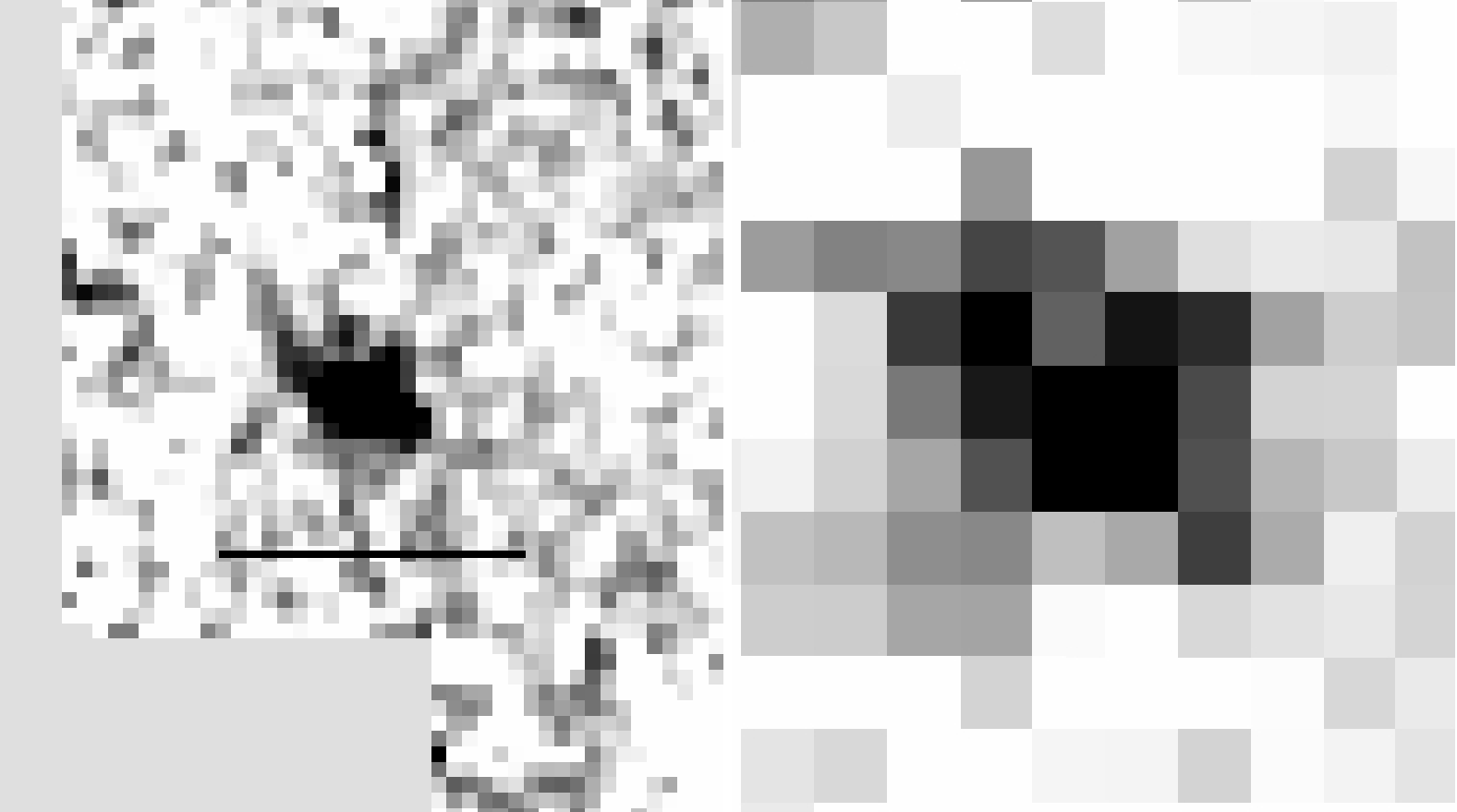} 
\includegraphics[width=4.4cm,bb=18 -180 592 718,height=5.4cm]{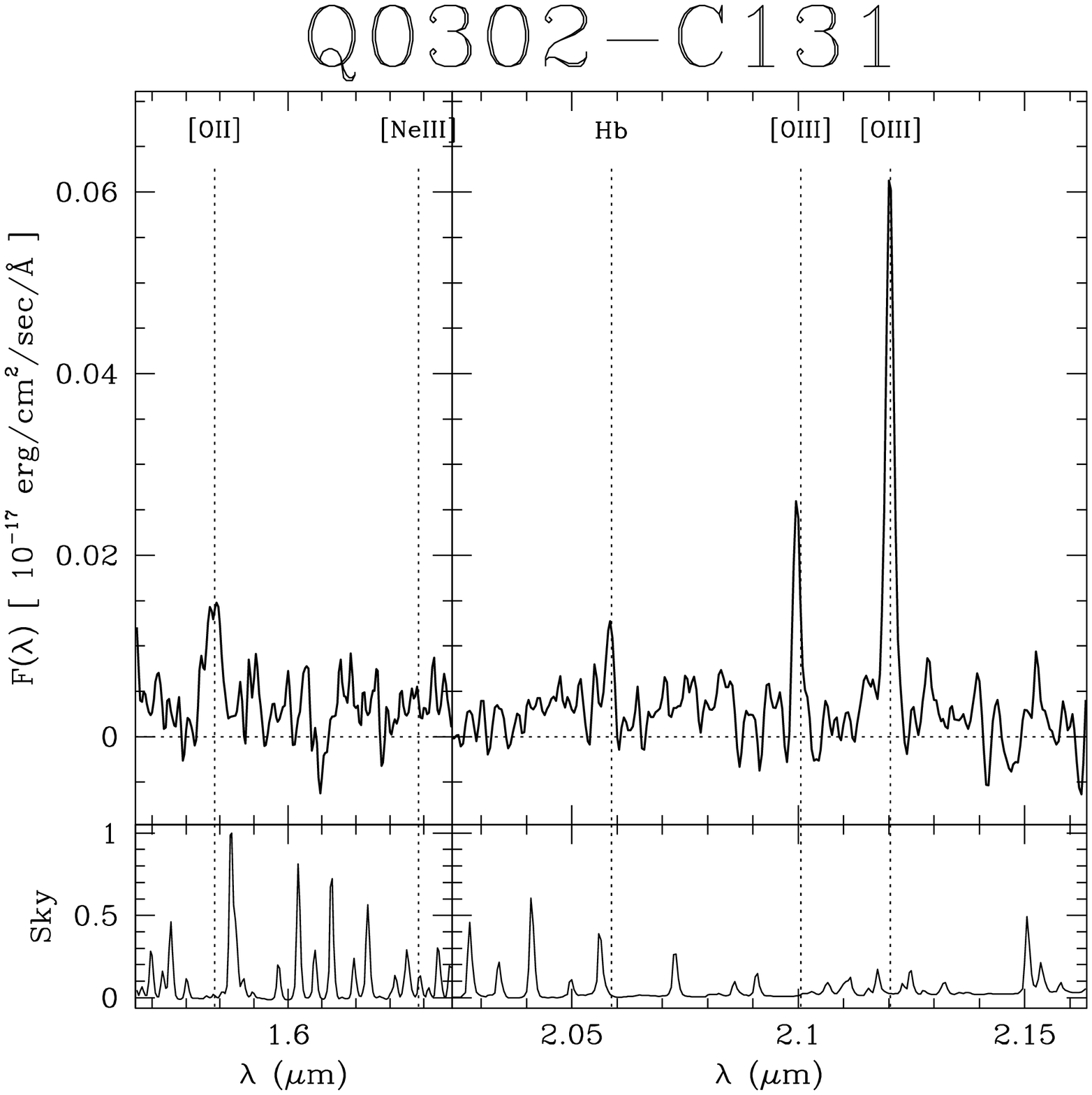} 
\\ \smallskip
\includegraphics[width=3.8cm,bb=0 -80 675 367]{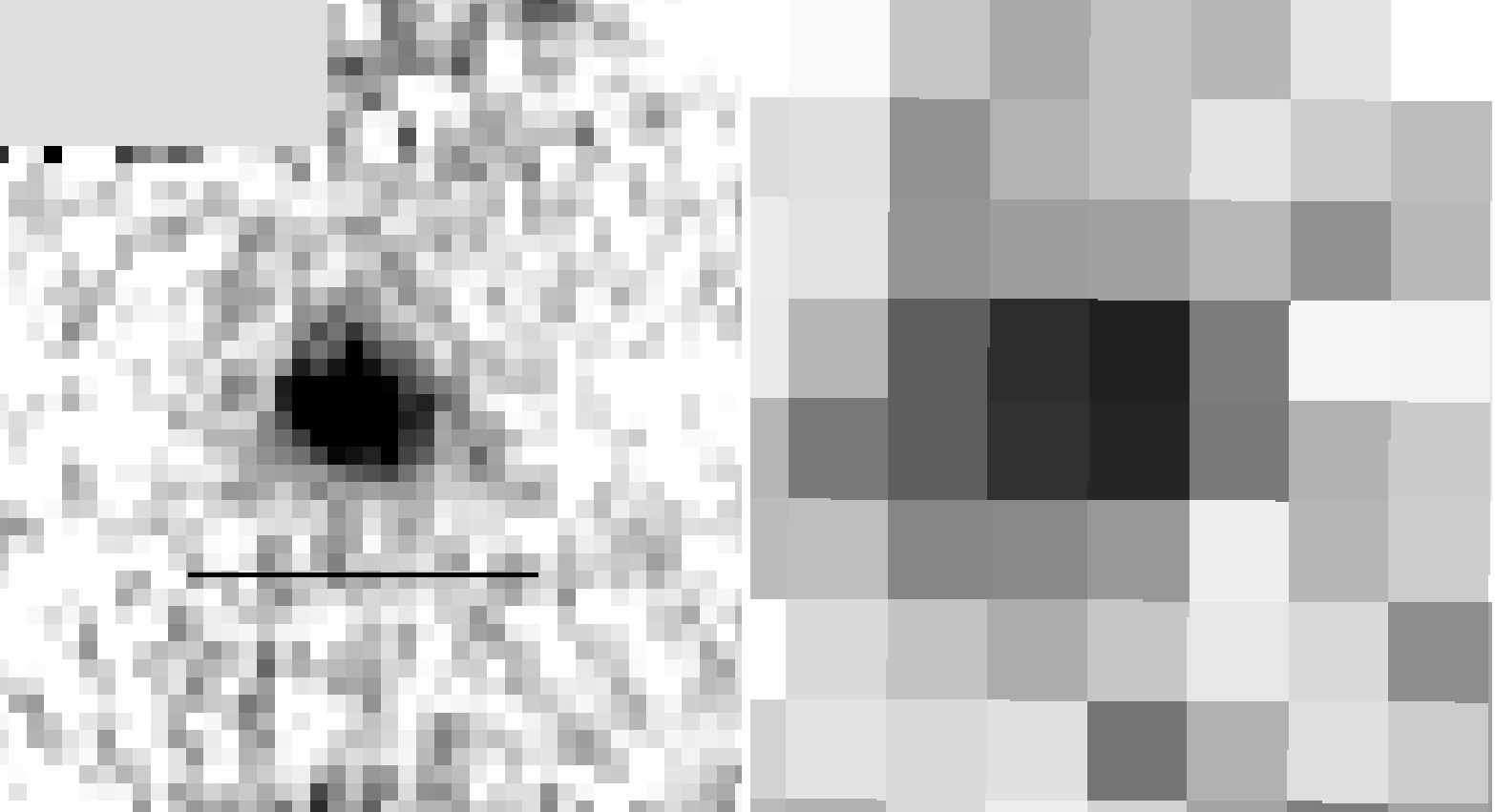} \hspace*{1.0mm}
\includegraphics[width=4.4cm,height=3.5cm]    {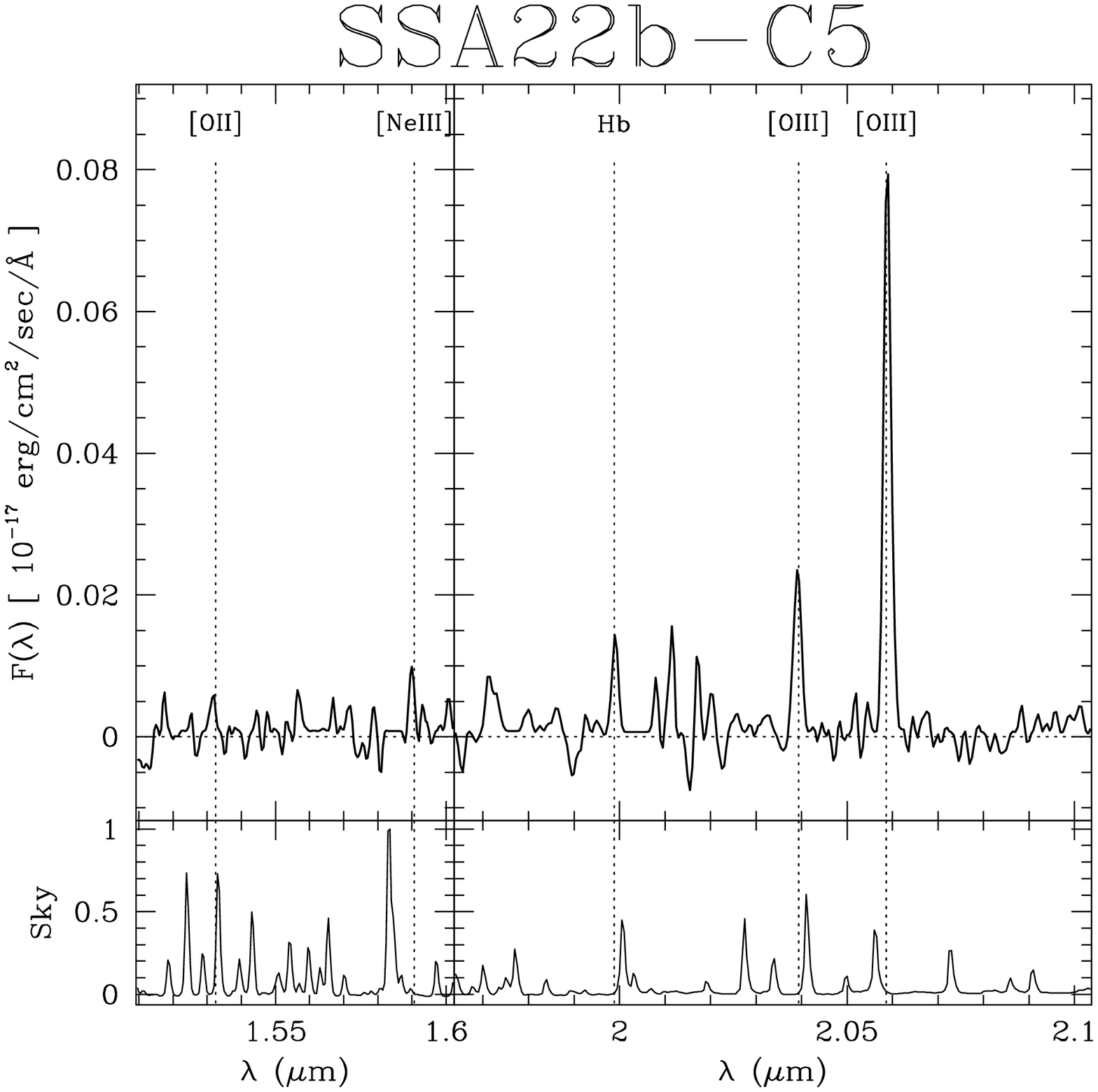}
\hspace*{2mm}
\includegraphics[width=3.8cm,bb=0 -100 898 414]{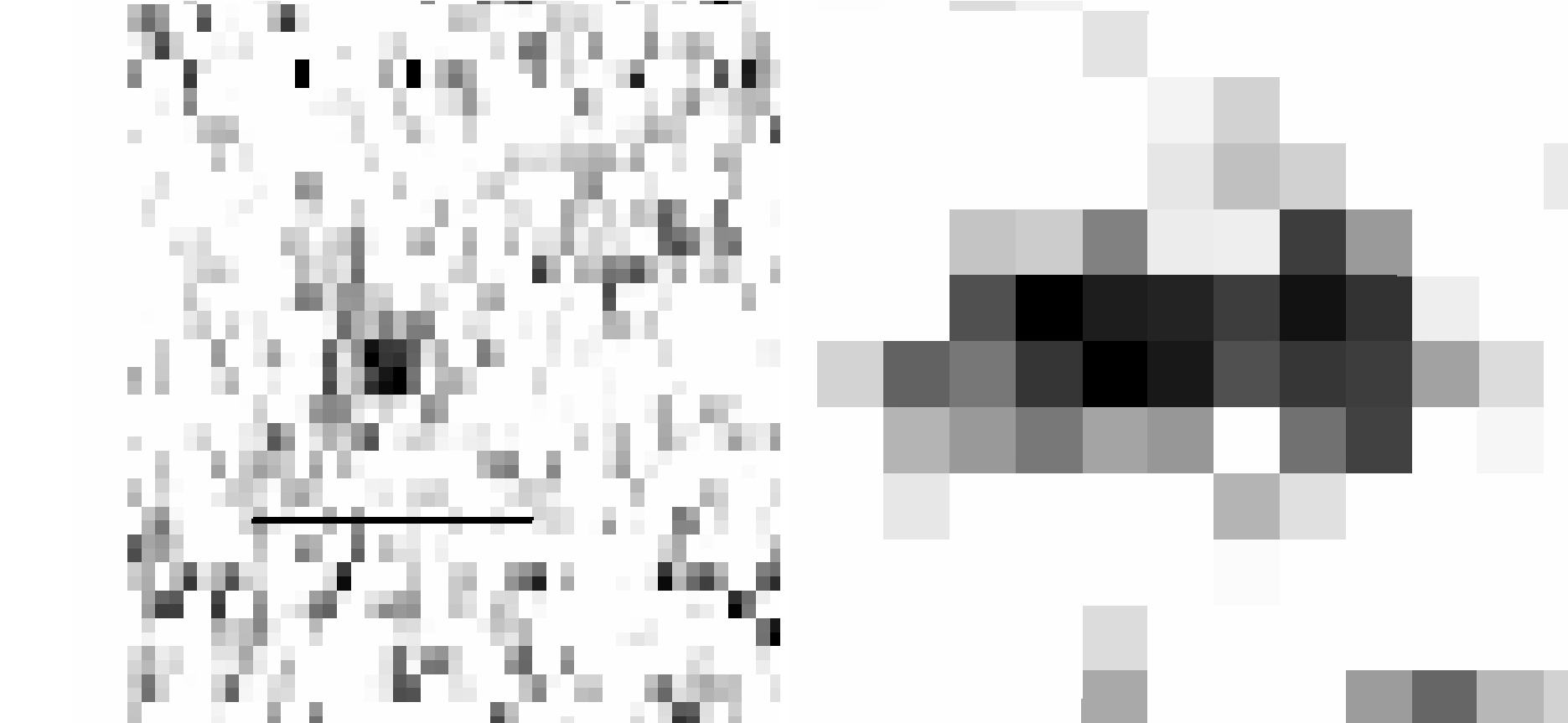} 
\includegraphics[width=4.4cm,height=3.5cm]     {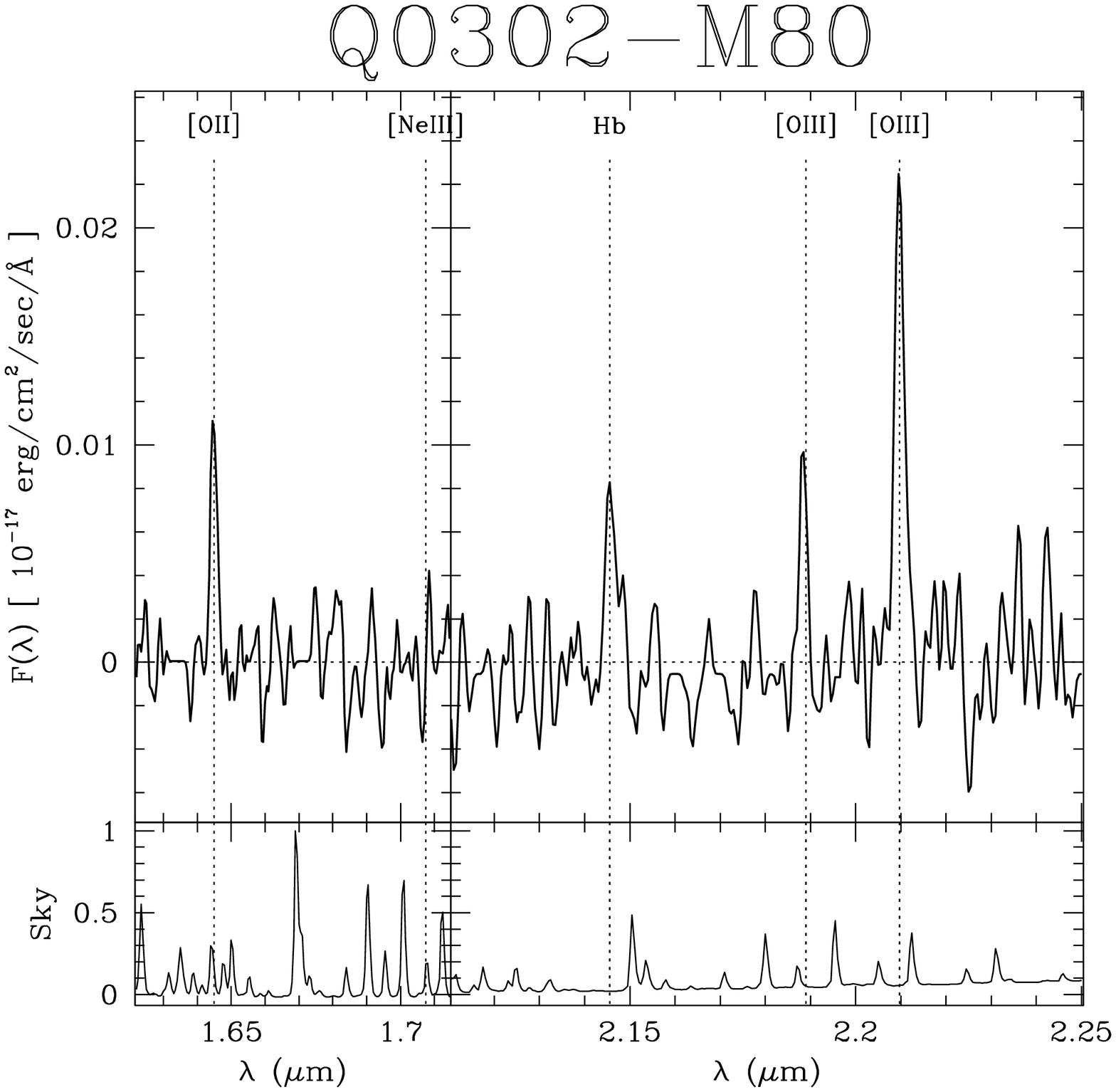} 
\\ \smallskip
\includegraphics[width=3.8cm,bb=0 -100 759 430]{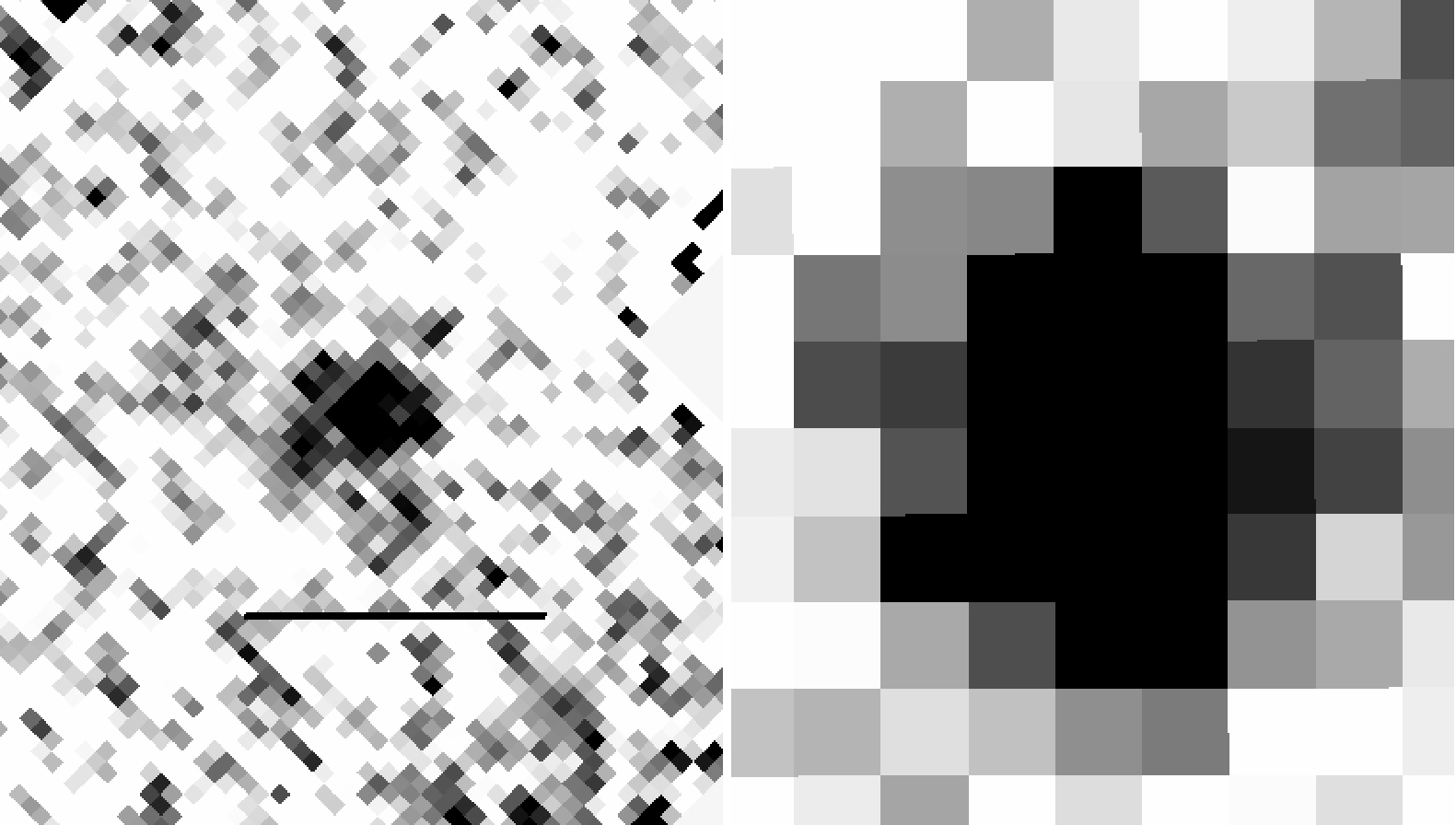}   \hspace*{2mm}
\includegraphics[width=4.4cm,height=3.5cm]     {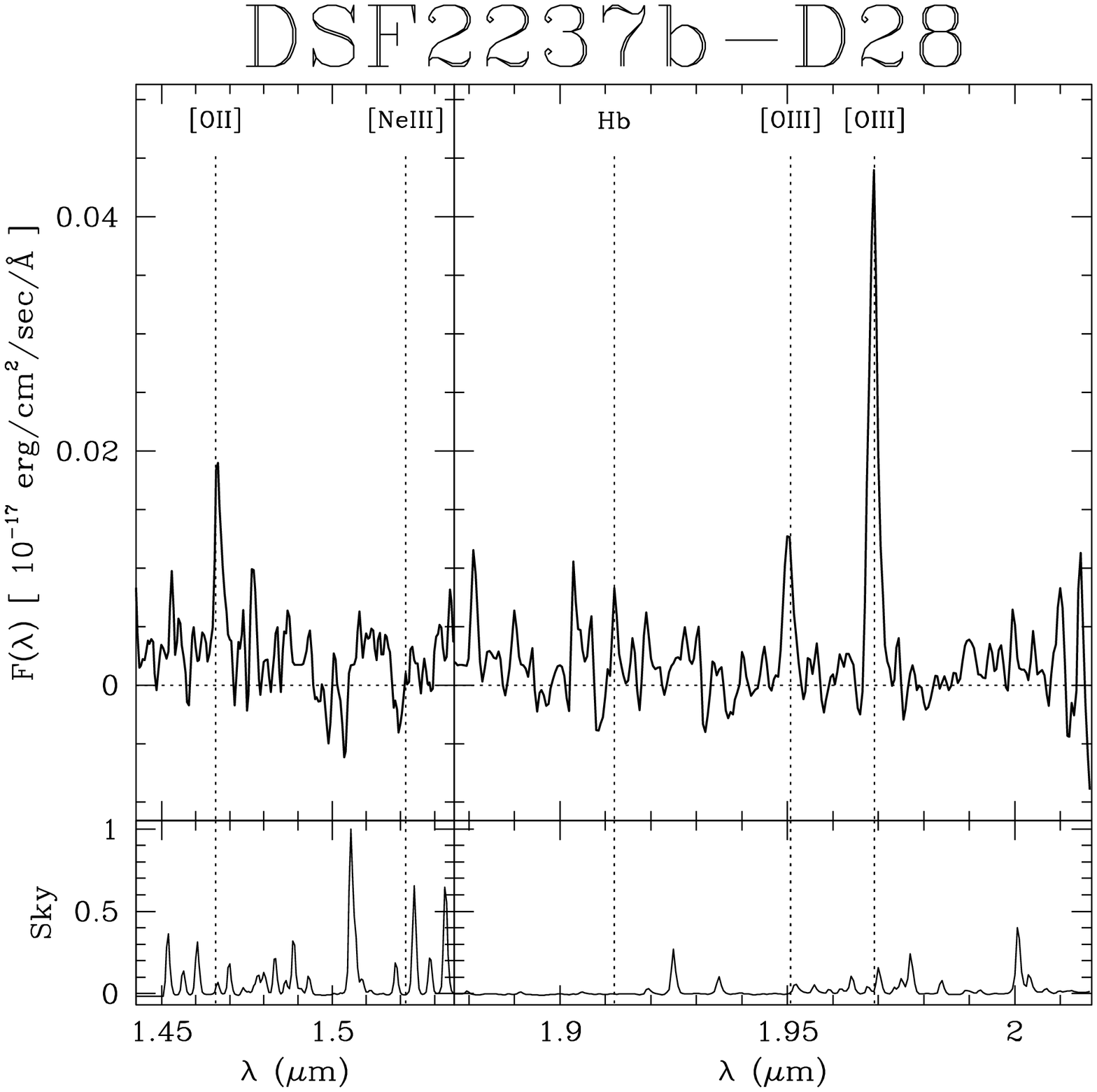} 
\hspace*{2mm}
\includegraphics[width=3.8cm,bb=0 -100 748 335]{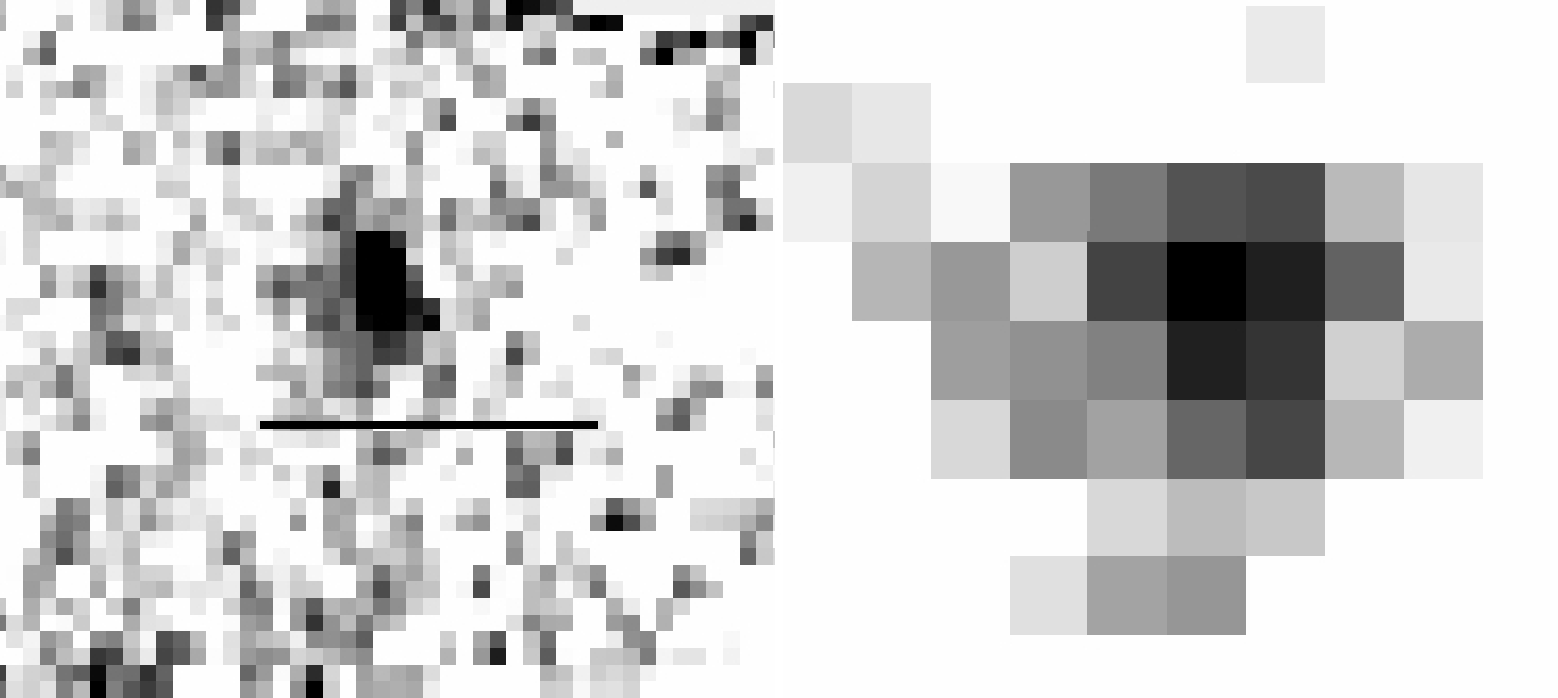} 
\includegraphics[width=4.4cm,height=3.5cm]      {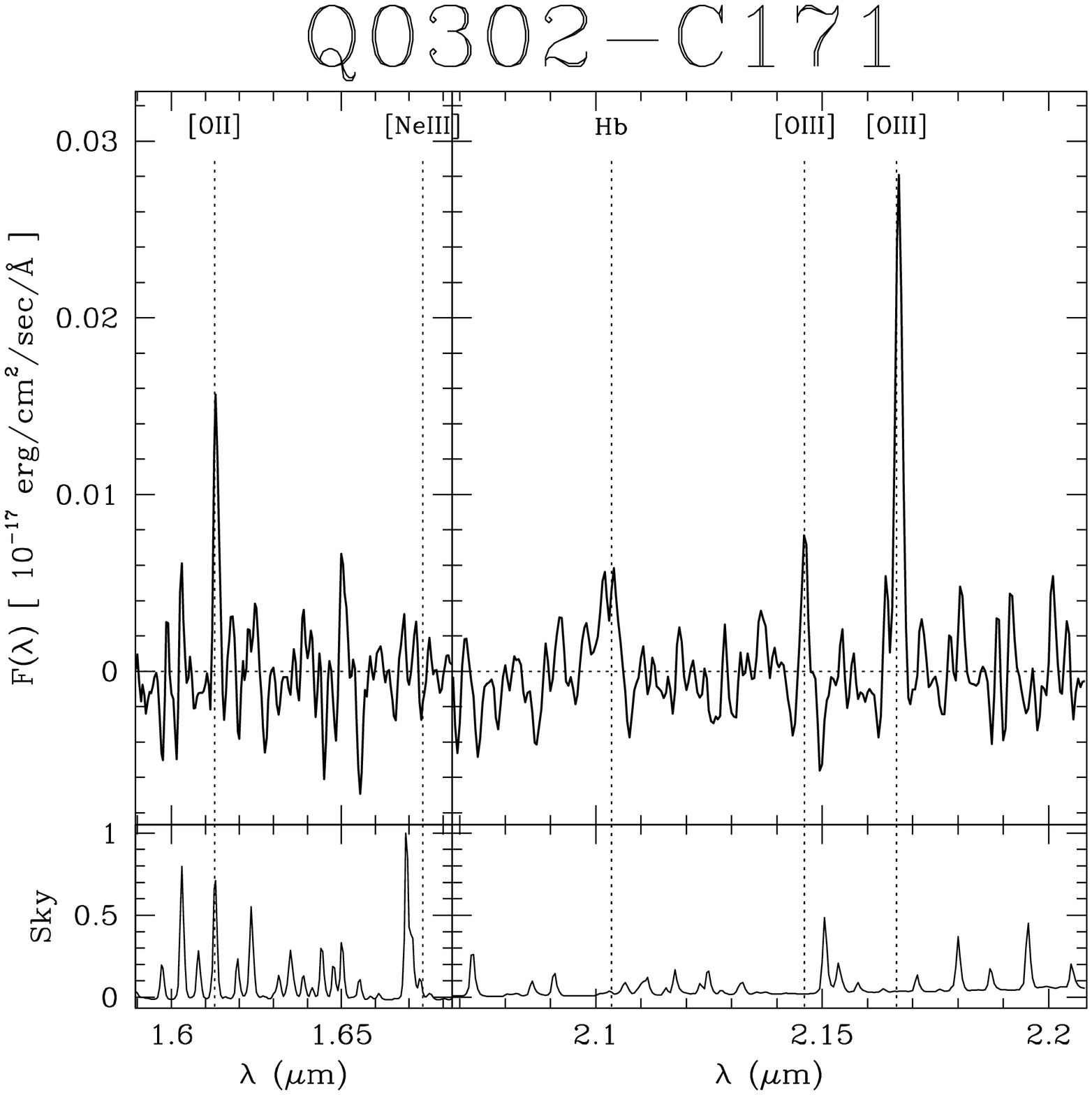} 
\caption{
Images and spectra of the targets. For each object we show, from left to right:
morphology in the emission line (\ha\ for MD19, \oiiib\ for the others),  
ground-based R-band image by \citet{Steidel03}, and galaxy spectrum
compared to the sky spectrum.
The horizontal bars in the line images are 1\arcsec\ long.
First column, from top to bottom:
SSA22a-C30, 
the interacting system composed by SSA22a-C6 (south, upper spectrum) 
 and SSA22a-M4 (north, lower spectrum), 
SSA22b-C5,  DSF2237b-D28.
Second column:
DSF2237b-MD19, Q0302-C131, Q0302-M80 and Q0302-C171.
}
\label{fig:images}
\end{figure*}

%-------------------------------------------------------------------
\subsection{Morphologies and sizes}

Fig. \ref{fig:images} shows the images of all the targets.
Most of the galaxies show a sharp peak of line emission and some extended structure.
In several cases, the objects appear to be composed by 
several clumps of emission. The distance between clumps is usually 
comparable to the dimension of the objects as revealed by ground-based
imaging. In a few cases the distribution of the secondary peaks follow the 
continuum light, while in some cases we detected faint ``companions'' 
that are not detected in broad-band.
VLT has a small pointing error, of the order of 0.2\arcsec,
but, compared to our resolution, this is enough to introduce a significant
uncertainty in the relative astrometry between UV and line images.
As a consequence, we cannot be sure where the line emission
originates with respect to the UV image.
Assuming that the main peak of line emission corresponds to the 
maximum of the UV light, 
the line-emission ``companions''
could be both secondary peak of star formation and line emission inside
a single formed galaxy, or the signature of an ongoing merging.
A detailed comparison with broad-band, rest-frame-UV morphology will be given 
on the basis of future HST images.\\

Images obtained with a NGS usually have a non-circular PSF (see, for example,
\citealt{Cresci05}). 
Increasing the distance of the scientific target from the NGS,
the PSF becomes larger. The variation in the radial direction, i.e., towards 
the NGS, is larger than in the tangential direction. As a result, when this effect
is dominant, the objects appear to be elongated toward the NGS.
The magnitude of the effect depends on the conditions of the atmosphere 
and on the distance of the target from the NGS, and is of the order of 
0.1\arcsec--0.3\arcsec.
In our images, elongation toward the NGS is seen only in one case, meaning that
in most cases the atmospheric conditions were good enough for AO observations 
and the isoplanatic angle in the K band
was not significantly smaller than the separation between the 
target and the guide star.

No bright point sources are present in our field-of-view, 
preventing us from measuring the final PSF on the combined data themselves.
To estimate the PSF we used the on-axis image of the NGS taken every hour
at the beginning of each OB. The on-axis PSF FWHM was degraded 
to take into account the distance of the target galaxy from the NGS.
For this we used the recipes in \cite{Cresci05}, using the values of the 
isoplanatic angle provided by the ESO database, and averaging over the 
different values obtained for the different OBs.
As the seeing was always comprised in the narrow range 0.6--1.0\arcsec\ FWHM, 
no large differences are seen in most of the OBs.
In all cases we obtained values of the PSF FWHM of the order of 0.2\arcsec
for the fields observed with the smallest spatial scale, and 0.3\arcsec
for the fields with the largest scale.
These values are in good agreement with the results of the simulations
provided by ESO, and were used in the following analysis.\\
 
Galaxy sizes have been estimated by fitting a Moffat function to the image
in the brightest line,
and the results have been deconvolved for the PSF.
The measured half-light radii are listed in Table~\ref{tab:massmet}. 
Large uncertainties on the resulting sizes are present,
both because images usually have low signal-to-noise ratio (SNR)
and because the PSF is not accurately known. Also, companions and
secondary peaks of emission are not considered in measuring
the dimensions. As this is also true for line fluxes, the derived metallicities
(section~\ref{sec:met}) and gas mass (section~\ref{sec:mgas}),
are consistently derived from the main peak only. \\

Since we are observing faint galaxies with high-spatial resolution,
surface brightness is probably our limiting factor.
The dimensions of the detected line-emitting regions are sometimes
significantly smaller
than the dimensions obtained in the UV by ground-based imaging, 
even taking into account  the larger PSF. 
They are also smaller than object sizes
obtained with SINFONI/VLT without the use of AO by the
Spectroscopic Imaging survey in the Near-infrared with SINFONI (SINS)
(\citealt{Forster06a}, although at $z$$\sim2.2$)
and AMAZE \citep{Maiolino08} collaborations. 
We interpret these results as evidence that our
detected sizes are indeed limited by surface brightness.
The average limiting surface brightness is 
$1.2\times10^{-16}$ erg/cm$^2$/sec/arcsec$^2$. 
Such a sensitivity can be translated into a limiting SFR density 
at $z$$\sim$3.1 detected with an emission line. 
For \hb, the limiting SFR density is $\sim$4/\msun/yr/kpc$^2$; for \oiiib, 
assuming 
an average ratio of \oiiib/\hb=4, the minimum detectable SFR density is
$\sim$1\msun/yr/kpc$^2$.
This is similar to the SFR density at $z$=2.2
in \cite{Erb06c}, and it is adequate to detect a significant fraction
of the local starbursts
\citep{Kennicutt98}. Nevertheless, the outer, low surface brightness
parts of these galaxies are probably lost in the noise, and morphologies could 
appear more compact than they actually are.

The measured half-ligh radii range between 0.7 and 2.4 kpc, with a median 
value of 1.36 kpc. 
These radii ar similar to those in the sample
of local starbursts in \cite{Lehnert96b}, who measure a median value 
of 1.70 kpc. 
Considering that our observations are less sensitive to the outer parts 
of the galaxies than \cite{Lehnert96a}, we conclude that LBGs tend to be
as extended as local starbursts.

%----------------------------------------------------------------------------
\subsection{The Stellar mass}
\label{sec:mass}
%-----------------------------------------------
\begin{table*}
\caption{Properties of the target galaxies}
\label{tab:massmet}
\begin{tabular}{lcccccccc}
\hline
\hline
  (1)      &    (2)                  &  (3)    &       (4)       &   (5)        &     (6)              & (7)           &  (8)                   &  (9)         \\
Object     & log(M*/\msun)           &A$_{V,c}$&     SFR$_{em}$  &  $r_{1/2}$   &  12+lg(O/H)          &lg(\mgas/\msun)&  f$_{g}$               & lg(y$_{eff})$\\
           &                         & range   &     (\msun/yr)  &   (kpc)      &                      &               &                        &              \\
\hline                                     
SSA22a-C30 & 10.33$^{+0.31}_{-0.38}$ & 0.0--0.8& 29$^{+81}_{-21}$&1.48$\pm$0.44 &8.16$^{+0.20}_{-0.60}$& 9.66$\pm$0.41 & 0.18$^{+0.26}_{-0.12}$ & $-2.63_{-0.7-0.2}^{+0.2+0.3}$\\   
SSA22a-C6  &  9.68$^{+0.15}_{-0.06}$ & 0.3--0.5& 23$^{+11}_{-8 }$&1.75$\pm$0.22 &7.95$^{+0.20}_{-0.50}$& 9.96$\pm$0.13 & 0.66$^{+0.08}_{-0.12}$ & $-2.22_{-0.4-0.2}^{+0.2+0.1}$\\
SSA22a-M4  &  9.41$^{+0.34}_{-0.13}$ & 0.0--0.8& 20$^{+40}_{-13}$&2.01$\pm$0.27 &8.12$^{+0.25}_{-0.45}$& 9.96$\pm$0.34 & 0.78$^{+0.12}_{-0.28}$ & $-1.84_{-0.4-0.4}^{+0.2+0.4}$\\
SSA22b-C5  &  8.96$^{+0.38}_{-0.22}$ & 0.0--0.5& 15$^{+15}_{-8 }$&1.28$\pm$0.38 &7.66$^{+0.20}_{-0.20}$& 9.64$\pm$0.21 & 0.83$^{+0.08}_{-0.19}$ & $-2.17_{-0.2-0.4}^{+0.2+0.3}$\\
DSF22-D28  &  9.78$^{+0.28}_{-0.29}$ & 0.0--0.6& 14$^{+18}_{-8 }$&1.50$\pm$0.48 &8.20$^{+0.10}_{-0.25}$& 9.62$\pm$0.25 & 0.61$^{+0.21}_{-0.20}$ & $-2.32_{-0.2-0.2}^{+0.1+0.3}$\\ 
DSF22-MD19 & 10.06$^{+0.33}_{-0.27}$ & 0.0--1.2& 39$^{+77}_{-26}$&2.00$\pm$0.55 &  --                  & 9.98$\pm$0.33 & 0.69$^{+0.25}_{-0.26}$ &     --                       \\
Q0302-C131 & 10.09$^{+0.10}_{-0.33}$ & 0.0--0.3& 10$^{+6 }_{-4 }$&1.27$\pm$0.37 &8.00$^{+0.25}_{-0.40}$& 9.57$\pm$0.14 & 0.42$^{+0.13}_{-0.07}$ & $-2.73_{-0.4-0.1}^{+0.2+0.2}$\\
Q0302-M80  & 10.07$^{+0.23}_{-0.19}$ & 0.0--0.6& 13$^{+17}_{-8 }$&0.75$\pm$0.24 &8.36$^{+0.15}_{-0.15}$& 9.54$\pm$0.25 & 0.36$^{+0.16}_{-0.12}$ & $-2.37_{-0.1-0.2}^{+0.1+0.2}$\\
Q0302-C171 & 10.06$^{+0.10}_{-0.28}$ & 0.0--0.2&  5$^{+2 }_{-2 }$&1.25$\pm$0.39 &8.14$^{+0.25}_{-0.45}$& 9.26$\pm$0.11 & 0.30$^{+0.08}_{-0.04}$ & $-2.71_{-0.4-0.1}^{+0.2+0.1}$\\
\hline                                   
\hline
\end{tabular}
Columns. (2-3) Stellar mass and dust extinction from SED fitting; 
(4): {\bf total SFR from emission lines}; 
(5): {\bf half-light radius of line emission}; 
(6): gas-phase metallicity
(7-8): gas mass and gas fraction; 
(9): effective yields; the first error is due to metallicity, the second one to gas fraction
\end{table*}

Stellar masses (\mstar) are derived by fitting the SED 
with {\it Hyperzmass} \citep{Pozzetti07},
a modified version of the 
photometric redshift code {\it Hyperz} \citep{Bolzonella00}. 
These fits also provide estimates of age, current SFR and dust
extinction of the dominant stellar population. The presence
of good IRAC photometry allows the determination of reliable 
\mstar\ as the SED is sampled up to the rest-frame J band. 
We have used the \cite{Bruzual03}
spectrophotometric models of galaxy evolution and
smooth exponentially decreasing SFHs, constraining the age to be
smaller than the Hubble time at galaxy redshift,
as detailed in \cite{Pozzetti07}.

The photometric stellar masses
have typical dispersions due to statistical uncertainties 
and degeneracies of the order of $0.2$ dex.
Metallicity has little influence on the SED, with uncertainties of $\sim$0.05 dex,
and we have set it to $Z$=0.2 \zsun, 
in agreement with what derived from the emission lines as described below. 
The addition of secondary bursts to
a continuous star formation history produces systematically higher 
(up to 40\% on average) stellar masses, while
population synthesis models with TP-AGB phase \citep{Maraston05} produce
a systematic shift of $\sim$0.1 dex towards lower \mstar. 
Finally, the uncertainty on the absolute value of the \mstar\ 
due to the assumptions on the IMF is within 
a factor of 2 for the typical IMFs usually adopted in the literature.
We have used the \cite{Chabrier03} IMF
with lower and upper cutoffs of 0.1 and 100 $M_{\odot}$.
The resulting stellar masses can be scaled to the standard Salpeter IMF by 
multiplying them by a factor of 1.7 \citep{Pozzetti07}.

The results are shown in Table~\ref{tab:massmet}. 
The listed errors are due to both statistical uncertainties and degeneracies.
Masses, expressed in log(\mstar/\msun),
range from 8.96 to 10.33, with a log average of 9.82. 
These values are consistent
with those derived by \cite{Shapley01} for their sample of LBGs at $z$$\sim$3,
averaging at 10.05 when converting their values to our IMF.
Lower redshift BM/BX galaxies in the $z$$\sim$2 sample of \cite{Erb06b} 
are also quite similar, with an average log(\mstar/\msun) of 10.22, within
$\sim$0.3 of our average value.

Galaxy age, current SFR and dust extinction on the continuum $A_{V,c}$
 cannot be individually well constrained
because they suffer of significant degeneracy. The errors associated to each of these
three quantities are large, and this is particularly important for 
extinction which is used to estimate the total SFR from emission lines.
The range of the allowed values of $A_{V,c}$ is 
shown in Table \ref{tab:massmet}.
Despite the large errors, a possible correlation is seen between stellar mass 
and age, i.e., 
less massive galaxies are best-fitted by younger templates.
The best-fitting age is of the order of $10^{7.5}$ yr for the less massive 
galaxies, and 
about $10^9$ yr for the most massive ones.

%----------------------------------------------------
\subsection{Photoionization conditions}

The values of SFR and metallicity derived in the next sections
are correct only if the optical line emission is dominated by star formation
and the conditions are not too different from those in the local universe.
Even if the observed line ratios are typical of starburst galaxies, several 
problems could be present.

The first concern is that the presence  of an AGN could alter line ratios 
and produce a spurious value of the inferred metallicity.
No evidence of AGN is seen in the emission line shape, where the width of the 
forbidden lines is consistent with those of the permitted lines. 
This excludes the
presence of a dominant broad-line AGN, but it does not help in excluding
a narrow-line AGN (see the discussion in sec.~\ref{sec:sample}).

The second concern is that significant differences in the conditions of 
star forming regions can be present, as LBGs at $z$$\sim$3 
have larger SFRs than most galaxies in the local universe. 
In particular, any evolution
of the photoionization conditions could mimic a change of metallicity.
The presence of evolution can be studied by comparing
the flux ratios [NII]/H$\alpha$ vs. [OIII]/H$\beta$,  and
several studies at lower redshift indicate that such an evolution 
actually exists
\citep{Shapley05a,Erb06a,Brinchmann08,Liu08}. 
Line ratios discrepant from the local relation are often 
associated with higher SFR
surface densities, interstellar pressures, and ionization parameters.

We have verified that all the observed line ratios are fully consistent
with excitation from hot stars, but we cannot study 
the presence of evolution because in most cases we cannot
observe \ha. Nevertheless, several authors 
\citep{Brinchmann08,Liu08}
conclude that the influence of different ionization conditions
on the derived metallicities is likely to be low, of the order of 0.1 dex.

%-----------------------------------------------
\begin{figure}
\centerline{\includegraphics[width=\columnwidth]{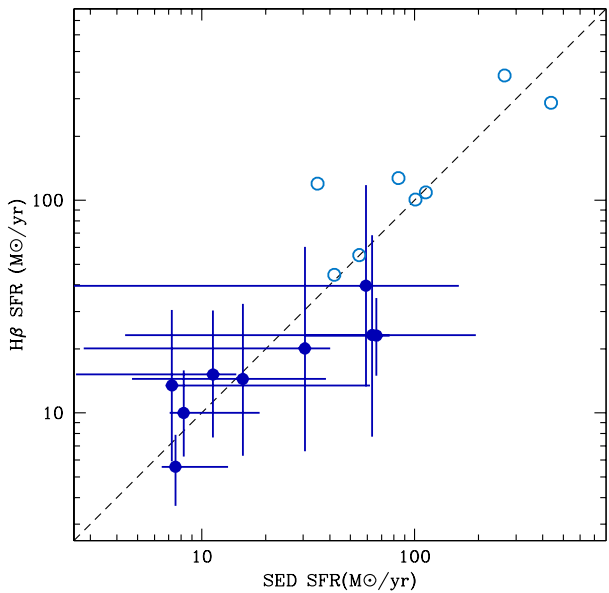} }
\caption{
Comparison between the SFRs derived from 
SED fitting and line emission. Solid and empty dots show LSD and AMAZE galaxies, 
respectively.  The dashed line shows equal SFRs. 
A good agreement is found between the two estimates of the SFR.
}
\label{fig:sfr1}
\end{figure}
%-----------------------------------------------

%-----------------------------------------------------------------------------
\subsection{SFR from optical lines and SED fitting}
\label{sec:sfr}

SFR can be derived from the \ha\ and \hb\ lines. 
Several conversion are possible, depending on the available data and 
physical condition of the galaxies. 
Similar to what is often done by many authors 
(e.g., \citealt{Erb06a,Forster-Schreiber09}),
we have used the ``classical'' 
conversion factor by \cite{Kennicutt98},
scaling down the results by a factor 1.7 \citep{Pozzetti07} 
to convert them to the \cite{Chabrier03} IMF. 
The \cite{Kennicutt98} conversion factor is based on the \hb\ flux corrected 
for dust extinction, and therefore requires a good knowledge of the amount of 
extinction $A_{V,el}$ suffered by the emission lines. 
\cite{Calzetti00}  found that this is proportional
to the extinction suffered by the continuum $A_{V,c}$, as measured by 
SED fitting,
with $A_{V,el}=A_{V,c}/0.44$ (see also \citealt{Forster-Schreiber09}),
and here we have used this assumption.
For DSF2237b-D28 the \hb\ flux is derived from
\oiiib\ and the expected line ratio for the value of metallicity 
measured in the next section.
The values of SFRs are listed in Table~\ref{tab:massmet} 
and plotted in Fig.~\ref{fig:sfr1}.

As explained in the previous section, large ranges of extinction are allowed by 
our SED fitting, therefore
the resulting SFRs have large errors, with typical uncertainties 
of one order of magnitude. This is common to most of the studies of 
high-redshift star-forming galaxies, even if sometimes the errors
on the SFR are not accurately discussed.
Unfortunately, other possible conversion factors that could provide smaller 
uncertainties cannot be applied to high-redshift starburt galaxies.
For example, several other estimates 
have been proposed by \cite{Argence08}, \cite{Moustakas06} and \cite{Weiner07},
based on a combination of two emission lines (such as \oii and \hb),
on lines and broad-band photometry, or 
line flux with no correction for dust extinction. These recipes are based on
the empirical correlation among mass, SFR, dust extinction, 
luminosity and metallicity
observed  in local galaxies, and it is not possible to apply them 
for galaxies with very different properties.

The SFR derived from the emission lines can be compared with 
the value resulting from the fit of the SED (see Fig.~\ref{fig:sfr1}). 
Large uncertainties
are present due to the errors on \hb\ flux, the
uncertainties in dust extinction correction for both estimates of SFR,
the intrinsic spread of both calibrations.
Nevertheless
a good agreement is seen between the two estimators of the SFR.

%=========================================================================
\subsection{Metallicity}
\label{sec:met}

Gas-phase metallicities were derived by comparing the observed line ratios 
with the calibrations in \cite{Nagao06} and \cite{Maiolino08}.
Metallicities are derived by a simultaneous fit of all the available line 
ratios, as explained in \cite{Maiolino08}. Both the uncertainties in the
observed line ratios and the spread of the calibration were considered, and in
some cases the latter is the dominant contribution to the metallicity error.
In practice, the result is dominated
by the R23 indicator or, similarly, 
by the \oiiib/\hb\ ratio, 
while the \oiiib/\oii\ ratio is used to discriminate between the two
possible branches at low- and high-metallicity. 
The [NeIII]3869/[OII]3727 line ratio is
also a sensitive indicator, when both lines are detected. 
The uncertainties on metallicity are of the order of 0.2--0.3 dex, and
tend to be larger when the metallicity has values 12+log(O/H)$\sim$8, when both
R23 and \oiii/\hb\ have a maximum. In this case a small uncertainty in the line
ratio produces a large uncertainty in metallicity.
The results are listed in Table~\ref{tab:massmet}.
LSD galaxies show metallicities about 10\%--50\% solar, similar to the values
found by \cite{Pettini01} for a different sample of LBGs at $z$$\sim$3.

Dust extinction is not strongly affecting the results as the only
extinction-sensitive line ratio is \oiiib/\oii. 
This effect can be clearly seen in Figure~6 of \cite{Maiolino08}, 
showing that metallicity is barely dependent on extinction and, as a consequence,
the observed line ratios leave extinction basically unconstrained.
For this reason we have chosen to limit the extinction to be within 
the range allowed by the SED fitting. Leaving extinction totally 
unconstrained does not affect the best-fitting value of metallicity
but only the range of confidence.

%----------------------------------------------------
\begin{figure*}
% \vspace*{-2.0 cm}
\begin{center}
\includegraphics[width=10cm,angle=-90]{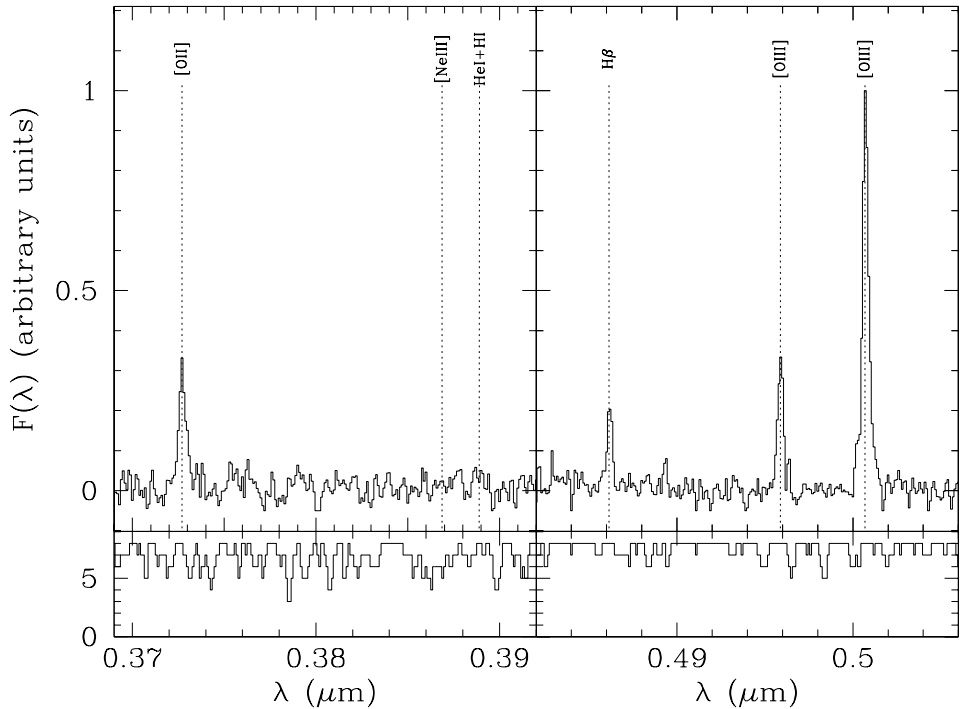} 
\caption{
Composite spectrum obtained by averaging the spectra of the 8 targets 
covering the spectral range between \oii\ and \oiiib.
The lower panel shows the number of spectra
used for each pixel, as the pixels interested by bright sky lines were not
included. The positions of the undetected lines \neiii\ and \hei\ 
are shown for reference.
}
\label{fig:composite}
\end{center}
\end{figure*}
%----------------------------------------------------

%=========================================================================
\section{Notes on individual objects}
\label{sec:objects}

\subsection{SSA22a-C30}
This is the galaxy with the largest stellar mass.
While \hb\ and \oiiib\ are in clean parts of the atmospheric spectrum, \oii\ and
\oiiia\ are affected by OH lines.
\oiiib\ emission is characterized by the presence of a 
secondary peak, about 1\arcsec\ apart from the main one. Other lower-luminosity
peaks can be present in between, although with lower significance. 
As a result, the object has a complex, very disturbed morphology.  
In the ground-based UV image, this galaxy appears less disturbed.
It is elongated as the line-emitting blobs and can comprise them all.

\subsection{SSA22a-C6 and SSA22a-M4}
This is a pair of interacting galaxies at a projected distance of about 
2\arcsec\ and with velocity difference of about 300 km/sec.
All lines, except \neiii, fall far from the OH atmospheric lines, allowing 
for a good measurement of the metallicity.
C6, the southernmost object, is the brightest member both in UV and in 
line emission. It has an elongated structure in the UV, and also the line 
image shows some secondary peaks  coincident with the UV elongation. 
M4 is compact in both UV and line images.

\subsection{SSA22b-C5}
This is the object with the smallest stellar mass, the only one 
with log(\mstar/\msun)$<$9.
While \oiii\ are well detected, \oii\ is hidden by a sky line.
A marginally significant
structure is detected near the expected position of the \neiii\ line, providing
a value of the redshift ($z$=3.116) slightly larger that the value from the
\oiiib\ line ($z$=3.112). This could be a real detection of the \neiii\ line, but
both the low statistical significance and the small velocity offset
make it quite uncertain. Here we assume that this is not a real line and 
do not use it to derive metallicity.
The galaxy shape is regular and almost circular, with no significant structure
around the main core.

\subsection{DSF22b-D28}
The line image is characterized by a compact nucleus surrounded 
by an extended faint emission, with a size much more compact than in 
the broad-band image. 
Its emission lines are quite faint and \hb\ is not detected.

\subsection{DSF22b-MD19} 
This is the galaxy of our sample with the lowest redshift, 
$z$=2.616, and it is the 
only one where \ha\ can be observed. This line falls in the K band at 2.37\mic,
where the thermal contribution to the background is significant. 
Nevertheless, it is very well detected. The nearby [NII] lines are 
not detected and we can put a limit to their flux ratio with 
\ha\ of [NII]6583/\ha$<$0.3. Such a value excludes
that the ionization could be dominated by an AGN.
The \oii\ line is outside the observed range and the \oiiib\ line is 
barely detected as it is coincident with a sky line. As a consequence 
we cannot measure the metallicity of this object. 
\hb\ is well detected in a clean part of the spectrum, and the 
\ha/\hb\ ratio is 
4.9$\pm$1.5, significantly above the case-B value of 2.85. 
Assuming that this difference is due to a screen of dust following the
\cite{Cardelli89} extinction law, we derive $A_{V,el}=1.7$. 
This is consistent with what is derived by the spectral fitting, $A_{V,c}$=1.1.
MD19 is very extended both in UV and in \ha.
The stellar continuum of this object is clearly detected and 
appears to be displaced from the peak of the lines.
\subsection{Q0302-C131}
This galaxy is elongated both in line and UV images, with a shape reminding 
a classical edge-on disk. The position angle of the elongation 
in the line image is fully consistent with that in the UV image.
Most of the lines are in a clean part of the spectrum, only the \neiii\ line
is hidden by a sky line.

\subsection{Q0302-M80}
In the broad-band image this galaxy is very extended, with an elongated 
structure. In contrast, the line image appears very compact and regular.
Both \oii\ and \neiii\ fall near bright sky lines and cannot be detected.

\subsection{Q0302-C171}
This is the object with the faintest \oiiib. 
While most of the main emission lines are not strongly affected by sky lines, 
the flux of \oii\ is very
uncertain because the line falls near a bright sky line at 1.613\mic.

\subsection{Composite spectrum}
The average spectral properties of the LSD galaxies can be better shown by
constructing an average spectrum. Also, fainter spectral features, below 
detection in any single spectrum, could be revealed by summing up all 
the spectra.
Fig.~\ref{fig:composite} shows the average,
composite spectrum of the 8 LSD sources whose spectra cover
the \oii--\oiiib\ region.
This spectrum was obtained as an average of the original ones without
any weighting, and
reveals the average properties of the LSD sample as if they were 
observed simultaneously. 
Large \oiiib/\hb\ and \oiiib/\oii\ ratios
relative to local galaxies are shown, and this is related to metallicity 
as explained in section~\ref{sec:met}.
The corresponding average mass is log(\mstar/\msun)=9.80.

%=========================================================================
\section{Analysis}

In the previous sections we have measured stellar mass, SFR, metallicity and size
of our sample of LBGs. In this section we intend to use these parameters to derive
information on the history and evolutionary state of these galaxies.
We will show that mass-metallicity relation, gas masses and effective yields
can provide important clues on the physical processes dominating the evolution 
of these objects and on the relation between these starburst and similar objects 
at lower redshifts.

%----------------------------------------------------
\begin{figure*}
\centerline{\includegraphics[width=13cm]{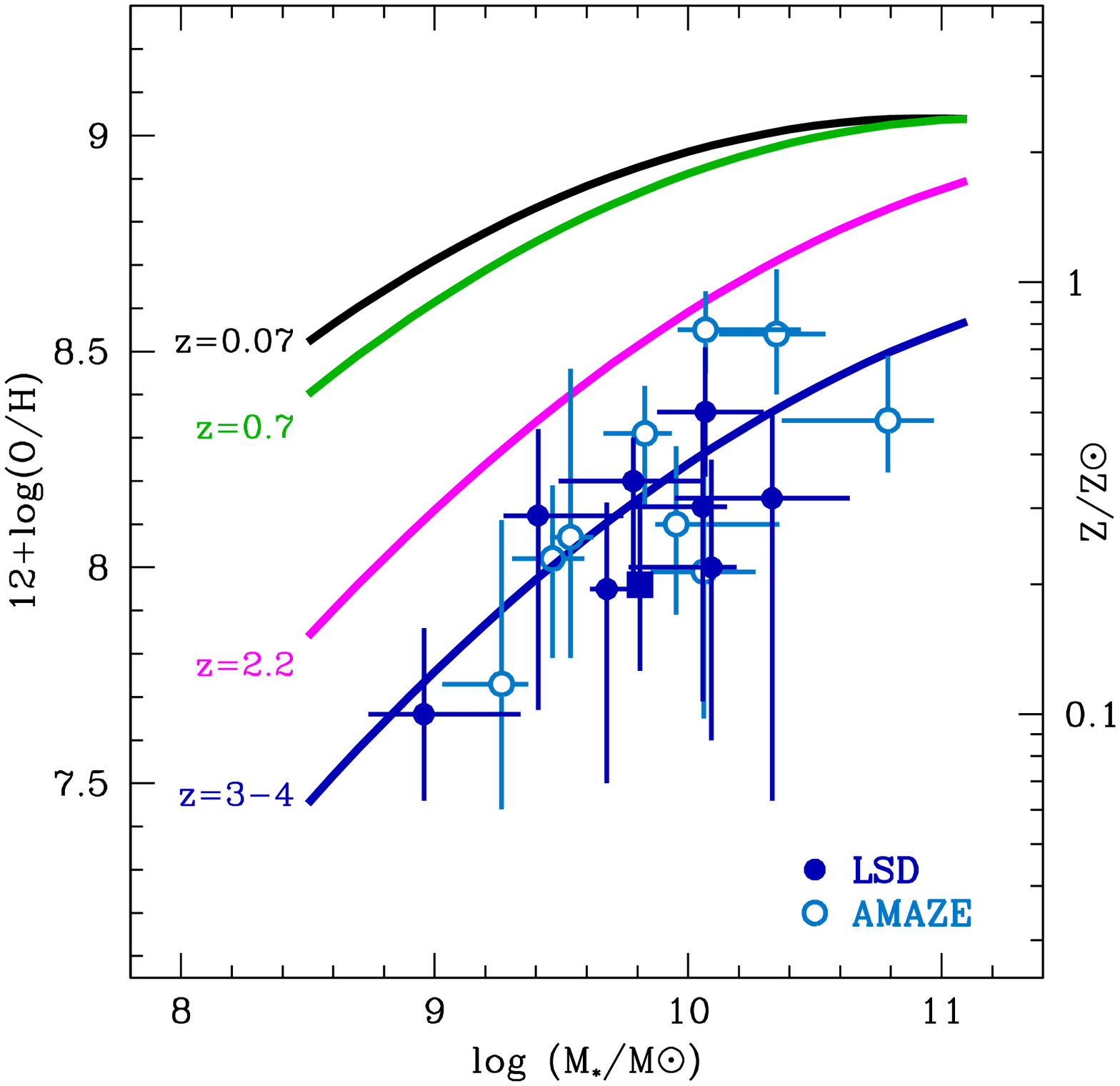} }
\caption{
Evolution of the mass-metallicity relation from $z$=0.07 \citep{Kewley08},
to $z$=0.7 \citep{Savaglio05}, $z$=2.2 \citep{Erb06a} and $z$=3--4 (AMAZE+LSD).
All data have been calibrated to the same metallicity scale and IMF 
\citep{Chabrier03} in order to make 
all the different results directly comparable.
Turquoise empty dots show the AMAZE galaxies, blue solid dots the LSD galaxies. 
The solid square shows the ``average'' LSD galaxy, having average mass and 
composite spectrum (see Figure~\ref{fig:composite}).
The lines show quadratic fits to the data, as described in the text.
}
\label{fig:massmet}
\end{figure*}
%----------------------------------------------------------------

%------------------------------------------------------------------------------
\subsection{The Mass-Metallicity relation}
\label{sec:massmet}

Figure~\ref{fig:massmet} shows 
the stellar mass-metallicity relation at $z$$\sim$3.1,
compared to the same relation as measured at lower redshifts.
All the presented data have been scaled to a \cite{Chabrier03} IMF
and use the same metallicity calibration.
A strong, monotonic evolution of metallicity
can be seen, i.e., galaxies at $z$$\sim$3.1 have metallicities 
$\sim$6 times lower than galaxies of similar stellar mass in the local 
universe. 
It is worth noticing that this is not the evolution of individual galaxies, 
as discussed in \cite{Maiolino08}, but this is the evolution of the
average metallicity of the galaxies contributing to a significant fraction of
the star-formation activity at their redshifts.
The observed evolution implies that  galaxies with relatively high stellar masses 
(log(M/\msun)=9--11) and low metallicity are already in place at $z>$3,
and this can be used to put strong constraints
on the processes dominating galaxy formation. 

While stellar mass, based on integrated photometry, is representative of the full
galaxy, metallicity is possibly dominated by the central, brightest
regions. The presence of metallicity gradients could have some influence on
the observed mass-metallicity relation. These aperture effects are present at
any redshift: even at $z$$\sim$0, galaxy spectra from SDSS refer to
the central few arcsecs of the galaxies. In most models, the central brightest
part of the galaxies are also the most metal rich, therefore the use of
total metallicities for our LSD galaxies is expected to produce an 
even larger evolution.

The effect of ``downsizing'' \citep{Cowie96} on chemical enrichment
is expected to produce differential evolution related to stellar mass.
Stronger evolution for low-mass galaxies is observed from $z$=0 to $z$=2.2
(see Figure~\ref{fig:massmet}).
The observed spread of the distribution and the uncertainties on the 
single points still make it impossible to see if such an effect
is already in place between  $z$=2.2 and $z$=3--4.
Constraints on this effect can be derived when the full AMAZE data sample
will be presented.

Using the same representation as in \cite{Maiolino08}, we fit
the evolution of the mass-metallicity relation with a second order polynomial:
$$Z=A \left[\rm{log}(M_*)-\rm{log}(M_0)\right]^2+K_0$$
where A=--0.0864 and $M_0$ and $K_0$ are the free parameters of the fit.
By using the LSD and AMAZE galaxies, we derive log(M$_0$)=12.28 and $K_0=8.69$.
The values of M$_0$ and $K_0$ for the samples at lower redshifts can be found
in \cite{Maiolino08}, and can be converted to the present system 
by subtracting log(1.7) to M$_0$.

\smallskip

Several published models of galaxy formation
(e.g., \citealt{Derossi07,Kobayashi07}) 
cannot account for such a strong evolution. The physical reason 
for this can be due to some inappropriate assumption,
for example about feedback processes or merging history. 
When taken at face value, 
some other models (e.g., \citealt{Brooks07,Tornatore07}) provide a better 
match with the observations, but a meaningful comparison can only
be obtained by taking into account all the selection effects and
observational biases, and by comparing not only stellar mass and metallicity
but also all the other relevant parameters, such as dynamical mass, angular 
momentum, gas fraction, SFR, morphology and size (see, for example, \citealt{Calura09}).

%-----------------------------------------------
\begin{figure*}
\centerline{\includegraphics[width=13cm]{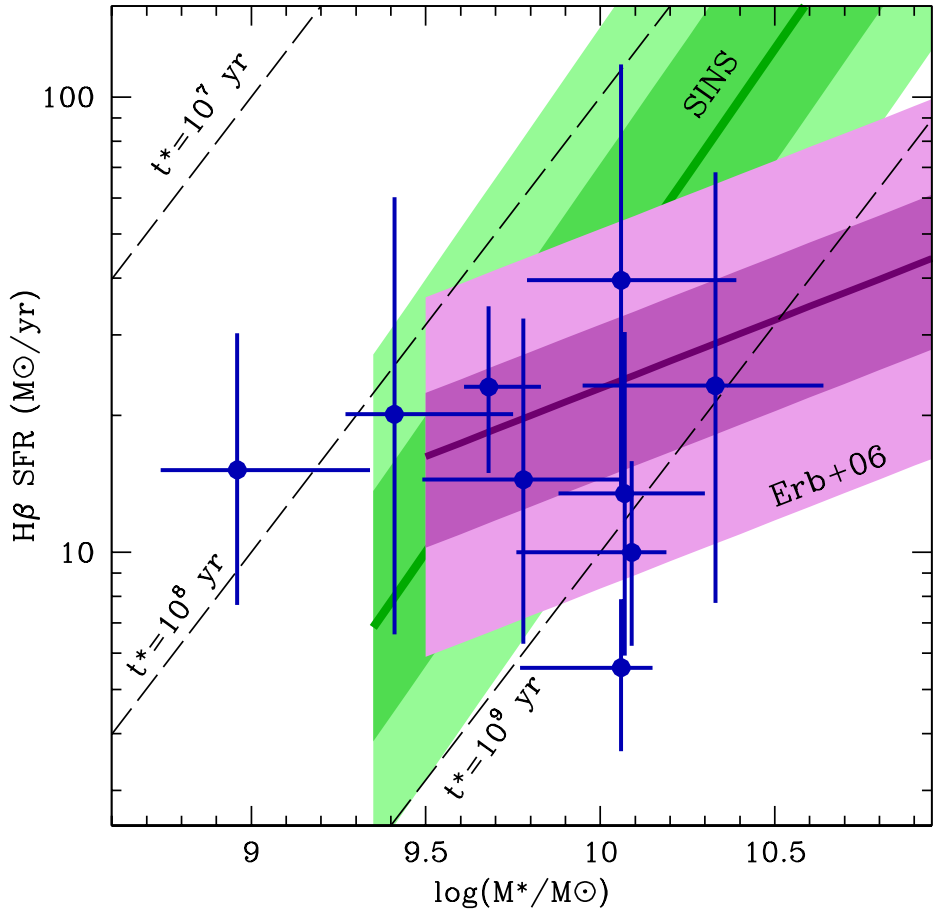} }
\caption{
SFR as a function of stellar mass for the galaxies in the 
LSD sample. This distribution is compared with two observed correlations at lower redshift 
($z$$\sim$2),
in magenta the UV-selected sample by Erb et al. (2006c), 
in green the SINS galaxies \citep{Forster-Schreiber09}.
The thick lines show the median of the correlations, while 
the color shaded regions are delimited by the 5\%, 25\%, 75\% and 95\% 
percentiles of each distribution.
Compared to the sample at $z$$\sim$2, LSD galaxies show a 
larger dispersion.
The long-dashed lines show
some values of star formation timescales, defined as $t_*$=\mstar/SFR.
LSD galaxies have SF timescales between $10^8$ and $10^9$ yr.
}
\label{fig:sfr2}
\end{figure*}
%-----------------------------------------------

In fact, it is important to emphasize that the galaxy samples used
for Figure~\ref{fig:massmet} change with redshift.
In the \cite{Tremonti04} work, the local SDSS galaxies under study constitute an 
almost complete census of the local star-forming galaxies, and the
derived mass-metallicity relation shows the average properties of the sample.
At high redshift, especially at $z>1$, a number of effects must be considered.
Only the most active galaxies are selected, and the observed metallicity refer 
to these objects. 
If, for example, less active, already formed and enriched galaxies are present 
at $z$=3, they would not be present in our sample;
the fraction of galaxies selected is likely to change with mass, and
this is expected to introduce some systematic effect with stellar mass;
the rest-frame UV selection misses extinguished galaxies, therefore an
increasing fraction of dusty, metal rich galaxies are excluded from 
the mass-metallicity relation at high metallicities; 
the UGR color-color selection for the LBGs excludes galaxies with red 
(G--R) colors due to the 
presence of older stellar populations. 
For example, \cite{Hayashi08} observed a sample of K-selected, 
star forming galaxies at $z$$\sim$2. 
They found a mass-metallicity relation which is similar to that in \cite{Erb06a} but 
with an offset of about 0.2 dex towards higher metallicities.
 The difference is likely to be due to
target selection. 
Evolution in the mass-metallicity plane can also be produced by different 
SF histories of individual galaxies:
if, for example, each galaxy experiences several bursts of star-formation,
galaxies detected during the first burst will show low stellar masses and low
metallicities, while galaxies with the same baryonic and total masses but
detected during later bursts will show higher metallicities and stellar
masses. This is not the case if galaxies have only one major burst, and all
the galaxies are selected during the same phase. 

All these effects are present at any redshift but their importance changes significantly
among the different samples. For this reasons, the quantitative 
interpretation of the observed 
evolution requires great care. A meaningful comparison between a sample of galaxies
caught during a special phase of their life at $z$=3,
with a more representative sample of galaxies at $z$=0 can only be done
when the selection effects are taken into account. 
This will be the subject of a future work.

%==========================================================================
\subsection{SFR and stellar mass}

Several studies of galaxies
have found a correlation between stellar mass \mstar\ and  SFR, i.e.,
more massive galaxies also have larger SFRs
(see, for example, \citealt{Schiminovich07} for a UV-selected sample,
\citealt{Elbaz07} for an optical selected sample, 
and \citealt{Daddi07} for a K-band selected sample).
In contrast, local Ultra-luminous Infrared Galaxies (ULIRG)
do not follow the same relation but lie above it \citep{Elbaz07}.
The slope of the correlation is not well constrained. 
While some authors \citep{Daddi07} find 
SFRs proportional to \mstar, i.e., specific SFR
(SSFR = SFR/\mstar) constant with mass, in most cases
a much weaker dependence of SFR on \mstar is found
\citep{Noeske07,Zheng07,Drory08}. 
Other authors find an evolution of the slope with redshift
\citep{Dunne08}, while in some samples
the SFR seems to be more directly related to stellar surface density 
rather than stellar mass \citep{Franx08}.

The existence of such a relation and its slope 
are strongly affected by sample selection. For example, color-based selections
tend to generate some proportionality between \mstar\ and SFR
because stronger SFRs are
needed to significantly affect the colors of more massive galaxies.
For 
UV-selected galaxies with SFRs estimated from the optical emission lines,
similar to LSD,
at $z$$\sim$2 \cite{Erb06a} have found 
a positive, albeit weak, correlation between \mstar\ and 
SFR by removing the galaxies which have dynamical mass much 
higher than the stellar mass.
In contrast, the SINS sample \citep{Forster-Schreiber09} 
shows a much stronger dependence of SFR on \mstar\, with slope consistent 
with one,
in agreement with the correlation found for the K-band 
selected sample in \cite{Daddi07}.

To produce such a correlation,
galaxies must spend a large portion of their life forming stars at
a characteristic fractional level, instead of forming most of their 
stellar mass in a few large bursts.
As a consequence, this correlation is often interpreted as an evidence 
that most of the star-formation activity
is not related to short, intense burst but rather is associated to a more 
continuos activity at a lower level. 
This is confirmed, at $z$$\sim$2.2, by the fact that the current SFR
in \cite{Erb06c} is similar to the past average.
This ongoing activity could be related to
cold accretion of gas followed by disk instabilities
\citep{Dekel06,Dekel09}. The regular morphologies of the galaxies dominating 
the star formation activity at intermediate redshifts \citep{Bell05,Wolf05}
and the (generally) regular dynamics observed at $z$$\sim$2
\citep{Forster06a,Genzel08,Shapiro08,Stark08}
support this interpretation. 

It is interesting to study if a correlation is present in our sample, and
Figure~\ref{fig:sfr2} shows the relation between \mstar\ and SFR in the 
LSD sample.
Large uncertainties are present, in 
particular on the SFR that is very sensitive to the poorly constrained 
amount of dust extinction. The LSD galaxies 
appear to have SFRs and \mstar\ similar to the lower-redshift 
galaxies in the samples mentioned above, but the small number of data points 
and the large intrinsic errors makes it impossible to obtain robust conclusions
about the existence of a correlation.

It is important to study if selection effects could 
hide a strong correlation between \mstar\ and SFR.
The rest-frame UV selection explains why no galaxy
below a given SFR threshold is selected. 
On the contrary, the lack of active, massive
could be due to several effects:
1- high-mass galaxies selected as LBGs with high SFRs could be rare
objects, and therefore none of them is included in this sample of 10 objects; 
2- it is possible that the dust content of the galaxies increases with age and 
metallicity, i.e., with stellar mass. In this case massive galaxies with large
SFRs could be too red or too faint in the UV to be selected as LBGs, and could
be present in other catalogs as sub-mm galaxies or ULIRGs. 
For example, this is what is observed by \cite{Nesvadba07} in the 
the archetypal sub-mm galaxy  SBS J14011+0252, which
is both metal rich, actively star-forming, and dusty;
3- the SF activity of each galaxy could be a decreasing function of time, with
a strong burst followed by a rapid decrease. In this case massive galaxies 
would also have low SFRs.

Albeit these limitations, our data do not support 
the existence of an almost direct proportionality 
(SFR $\sim$\mstar$^{1.0}$) between
SFR and \mstar\ similar to that in \cite{Daddi07} 
and \cite{Forster-Schreiber09}. 
In contrast, we find better agreement
with a flatter relation (SFR $\sim$\mstar$^{0.3}$) as found by \cite{Erb06a}.
This implies the presence of larger SSFRs in lower mass galaxies. 
As shown
by the dashed lines in Fig.~\ref{fig:sfr2},
the same effect can be expressed by 
the star-formation timescale, defined as the time $t_*$ to form the 
present amount of stars \mstar\ at the present level of SF, 
$t_*$=\mstar/SFR=1/SSFR.
In the LSD sample, $t_*$ changes systematically with \mstar: low mass galaxies 
have $t_*\sim 10^8$yr, while higher mass galaxies have $t_*\sim 10^9$yr.
This is an indication that, in relative terms,
the current episode of star formation is more important in lower-mass systems.
While high-mass galaxies have SFRs and SSFRs similar to what observed
at lower redshift, low-mass galaxies tend to be relatively more active.
This interpretation implies that less massive galaxies are also younger. 
An accurate test of this prediction is not easy, as it is difficult to measure the
age of high-redshift galaxies. Nevertheless, it is consistent with
the broad correlation between stellar mass and age described in 
sect.~\ref{sec:mass}.

%---------------------------------------------------------------------
\begin{figure}
\centerline{\includegraphics[width=7cm]{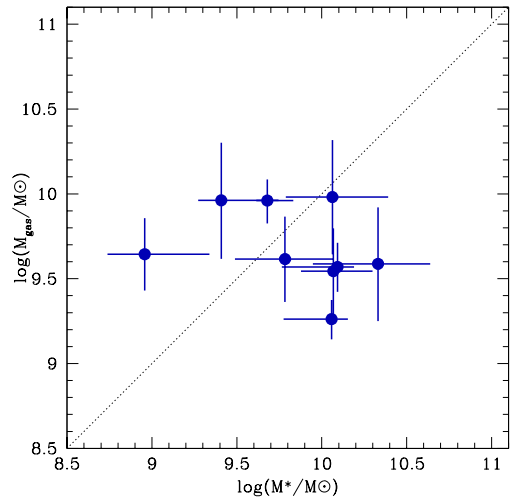} }
\caption{
Mass of gas as a function of the stellar mass for the LSD galaxies.
The dotted lines shows equal masses. All the galaxies have similar amount 
of gas, which is not related to stellar mass.
}
\label{fig:mgas}
\end{figure}

%=========================================================================
\subsection{The mass in gas}
\label{sec:mgas}

In local star-forming galaxies,
the surface density of star formation $\Sigma_{SFR}$
is related to the surface density of gas $\Sigma_{gas}$:
$$ \Sigma_{SFR}\propto\Sigma_{gas}^n$$
This is the Schmidt-Kennicutt law (see, for example, \citealt{Kennicutt98}),
which seems to be already in place at high redshifs
\citep{Bouche07}. 
It is interesting to note that downsizing is a natural consequence of the
Schmidt law and of the other scaling relations, 
as shown by \cite{Erb08}. 
If infalls and outflows can be
neglected on the timescale of a starburst, or if they are proportional to the
SFR itself, the Schmidt law produces a decreasing SFR 
whose initial, maximum value 
scales with the mass $M$ and the size $r$
as $M^{1.4}r^{-0.8}$, and the typical timescales as $M^{-0.4}r^{0.8}$.
If we assume that galaxies follow the same mass-size relation observed in the
local massive, early-type galaxies by \cite{Shen03}, 
we obtain that the typical timescale
decreases with mass as $M^{-0.1}$ and, correspondingly, the magnitude
of the initial starburst increases as $M^{1.1}$. 
The typical timescales computed by \cite{Erb08} 
are $\sim$0.3~Gyrs for $M=10^{12}$\msun, 
and $\sim$4~Gyr for $M=10^9$\msun.\\

The Schmidt-Kennicutt law is often used to 
estimate gas density and gas mass 
starting from the observed SFR and galaxy size
(for example, \citealt{Erb06a,Erb08}). 
The slope $n$ of the relation is still uncertain. 
The ``classic'' value of $n$ is 1.4, given by \cite{Kennicutt98}.
Star formation due to large-scale instabilities of the disk 
naturally produces $n$=1.5 \citep{Elmegreen02}, and such a value is supported
by more recent observations \citep{Kennicutt07} on small scales. 
In high redshift studies, \cite{Bouche07} found a slightly larger value,
$n$=1.7, a value which depends on the adopted value of the CO/H2 ratio.
Such a value could be more appropriate for our LBGs, nevertheless
for consistency with previous works 
we adopt the relation in
\cite{Kennicutt98}, deriving the following equations
for the gas surface density $\Sigma_{gas}$ and  the gas mass \mgas, valid when the
\cite{Chabrier03} IMF is used:

\begin{equation}
\Sigma_{gas}(M_\odot/{\rm pc^2})=254\ \left(\frac{\rm SFR}{\rm M_\odot/yr}\right)^{0.71}\ 
                                   \left(\frac{r}{\rm kpc}\right)^{-1.42}
\end{equation}

\begin{equation}
M_{gas}(M_\odot) = 798\ \left(\frac{\rm SFR}{\rm M_\odot/yr}\right)^{0.71}\left(\frac{r}{\rm kpc}\right)^{0.58}
\end{equation}

Using the values in \cite{Bouche07} would result in gas fractions about 40\% 
lower, which however do not affect the conclusions of this work.

The largest uncertainties in these equations are related to the 
half-light radius $r$, which has large errors, as explained above. 
This is important for $\Sigma_{gas}$, which has a strong dependence
on galaxy size. In contrast, the total gas mass \mgas\ is much less sensitive
to galaxy size as it depends on $r^{0.58}$ and, as a consequence, 
the uncertainties on this quantity are smaller.

The observed SFR densities are between 0.5 and 6\msun/yr/kpc$^2$,
similar to what observed both in the local universe \citep{Kennicutt98}
and at $z$=2.2 \citep{Erb06c}. 
The corresponding gas densities 
range between 300 and 2000 \msun/pc$^2$,
similar to the
values derived for LBGs and ULIRGs by \cite{Coppin07} and 
\cite{Tacconi06}.
As we are sampling the central, most active part of the galaxies, 
these numbers must be 
considered as the maximum surface densities of gas.

Figure~\ref{fig:mgas} shows the resulting \mgas\ as a function stellar mass.
It is evident that gas mass does not correlate with \mstar\ and that 
\mgas\ covers a much smaller range than \mstar.
While the largest galaxy has a stellar mass 24 times larger than the smallest,
the most gas-rich galaxy has only 5 times more gas than the poorest.
This result could be partly due to a combination of selection effects and 
spatial resolution. As all the galaxies have similar SFRs, the range in gas mass 
can be dominated by the range in intrinsic dimensions. 
Our resolution, 0.2\arcsec\ FWHM, corresponding to about 1.4 kpc, 
is similar to the intrinsic galaxy size. 
As a consequence, it is possible that we are overestimating the size,
and therefore the gas mass, of the smallest galaxies.\\

Interesting information on the physical properties of the LSD galaxies can 
be obtained by comparing the different timescales.
We have already introduced the {\em stellar} timescale $t_*$, which measures 
the time to create the
observed amount of stars at the current SFR.
The typical {\em dynamical} timescale is
$t_{dyn}=2\pi R^{3/2}/(GM)^{1/2}$, where $R$ is the typical radius and $M$ 
the typical total mass. 
Any starburst activity is expected to last at least $t_{dyn}$
\citep{Dekel06}.
The LSD galaxies have $t_{dyn}$ 
between 0.4 and 1.0$\times10^8$ yr.
This narrow range of values is not surprising, because
if size depends on mass as in local early-type galaxies, dynamical
times are almost independent of mass.
A third important timescale is the {\em gas exhaustion} time $t_{ex}$, i.e., 
the time to exhaust all the available gas at the current rate of SF, 
$t_{ex}=M_{gas}/SFR$. For the LSD galaxies, 
this spans a narrow range, between 3 and 6$\times10^8$ yr.

We have already seen that \mgas/\mstar, and therefore $t_*/t_{ex}$,
varies systematically with \mstar, meaning that the smallest galaxies are 
more gas rich
and are forming a larger fraction of stars. The ratio between $t_{dyn}$ and $t_{ex}$
ranges between 6 and 4, with no clear dependence on \mstar. This implies that
the galaxies are so gas rich that the current level of star formation 
can be sustained for several $t_{dyn}$. We cannot be sure that the SF activity will 
proceed for so long, but this is an indication that we are dealing with major
episodes of star formation in galaxies which have accreted large amounts of gas,
of the order of $10^{10}$\msun (see Fig.~\ref{fig:mgas}). In these galaxies
star formation follows mass assembly.
\\

%---------------------------------------------------------------------
\begin{figure}
\centerline{\includegraphics[width=\columnwidth]{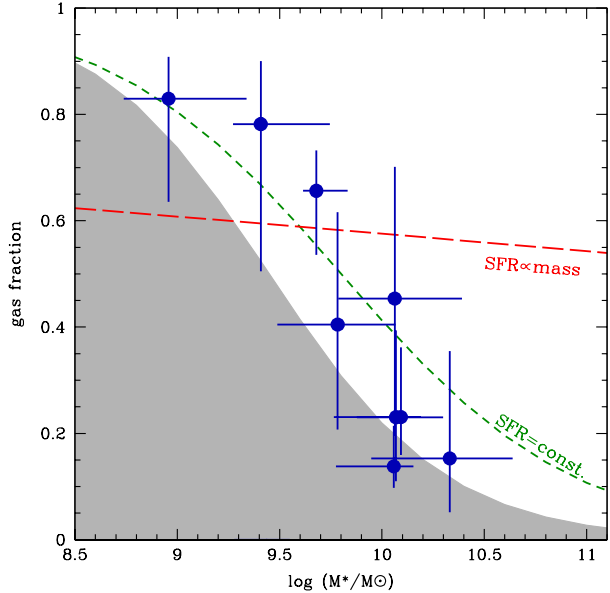} }
\caption{
Fraction of baryonic mass in gas as a function of stellar mass.
The blue solid dots are the LSD galaxies. 
The grey-shaded area represents galaxies that
cannot be included in our sample because the SFR is lower than our limit 
($\sim$15 \msun/yr).
The dashed lines show the expected dependence of gas fraction
on stellar mass for galaxies having SFR constant (green)
or proportional to the stellar mass (red).
}
\label{fig:gasden}
\end{figure}

%=========================================================================
\subsection{The fraction of gas}
\label{sec:fgas}

\smallskip

The gas mass \mgas\ can be compared with the total baryonic mass $M_b$
to estimate the fraction of mass in gas, an indication of the evolutionary 
stage of the galaxy. The results are shown in Figure~\ref{fig:gasden}, 
where the gas fraction is plotted vs. the stellar mass. 
It should be noted that stellar and gas masses are not measured in the same
apertures. When HST images for these objects are available 
(see sect.~\ref{sec:photom}), we will be able to quantify this effect and
apply a correction.
Despite the large uncertainties, a clear
correlation is seen between stellar mass and gas fraction. 
This is similar to what is observed at low-redshifts, where a tight 
correlation exists between gas fraction and rotation velocity 
\citep{Dalcanton07} 
or stellar mass \citep{Bell03a, Kannappan04}.
The same effect is also observed by \cite{Erb06a} and \cite{Reddy06}
in their samples at $z$$\sim$2.
The relation between stellar mass and gas fraction in local galaxies 
is often used to constrain the models of galaxy formation 
(e.g., \citealt{Somerville08}).

The observed correlation 
is at least partly due to the selection effects of our target sample, shown as
a grey-shaded area. 
As explained above, our sample is selected according to the SFR, while
stellar mass can assume a wide range of values.
The existence of this selection effect makes it impossible to test at $z$=3 the
results of several models of galaxy formation, 
such as \cite{Brooks07} and \cite{Mouchine08}.
These models predict that a strong correlation exists between \fgas\ and \mstar\ at any redshift, 
with more massive galaxies having a lower content in gas, very
similar to what is actually observed (see also \citealt{Calura08}).
In other words, the shaded region in Fig.~\ref{fig:gasden} 
could be empty for physical reasons.

The {\em expected} gas fraction as a function of stellar mass can be derived
once the dependence of dimensions and SFR on \mstar\ is known.
We compute this gas density using 
the mass-size relation for massive, late-type galaxies in \cite{Shen03},
$r\propto M^{0.4}$, 
and assuming that the SFR of the galaxies does not depend on stellar mass.
As size are expected to change slowly with mass, these results are 
largely independent of the evolution of the mass-size relation, 
which is largely unknown.
The result is plotted as a short-dashed, green line in Figure~\ref{fig:gasden}.
In contrast, assuming that the SFR is proportional to stellar mass
as observed in the local universe (see section~\ref{sec:sfr}), we derive
the long-dashed, red line in the same figure.
The normalization of these two lines can vary 
as a function of the adopted SFR and galaxy size, but the shape of the curves 
show that the gas fraction measured in our target galaxies is that 
expected for a SFR-limited sample.\\

%---------------------------------------------------------------------
\begin{figure}
\centerline{\includegraphics[width=\columnwidth]{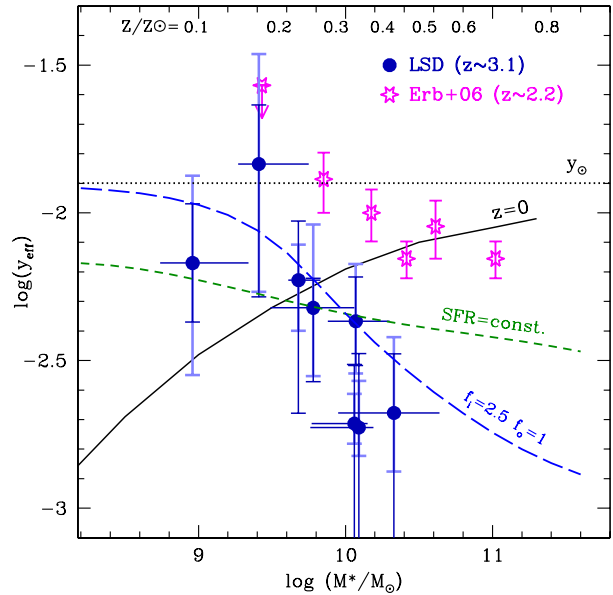} }
\caption{
Effective yields of the LSD sample (solid blue dots), as a function of
stellar mass, compared with the solar value (dotted line). 
Light and dark error bars show the contribution to the errors
on yields due to the uncertainties on gas fraction and metallicity,
respectively. 
The magenta stars are the results by \citealt{Erb06a} at $z$=2.2.
The black solid line are the local results by
Tremonti et al. (2004).
The green dashed line corresponds, as in
Figure~\ref{fig:gasden}, to galaxies with constant SFR.
The blue, long-dashed line is the model with infalls and outflows
proportional to the SFR described in the text.
The labels near the top of the figure show the values of the average
metallicity for galaxies of a given stellar mass as derived from
the mass-metallicity relation.
}
\label{fig:yields}
\end{figure}

%=========================================================================
\subsection{The effective yields}
\label{sec:yields}

Important hints on the physical processes shaping the mass-metallicity
relation can be obtained by 
considering the effective yields, i.e., the amount of metals produced 
and retained in the ISM per unit mass of formed stars
\citep{Garnett02,Dalcanton07}. 
In a closed-box model, metallicity $Z$ (i.e., the mass
in metals divided by the mass of the interstellar gas) can be written
as a simple function of the gas fraction \fgas:
\begin{equation}
Z=y\cdot ln(1/f_{gas})
\label{eq:yields}
\end{equation}
where $y$ is the true stellar yield, i.e., the ratio between the 
amount of metals produced and returned to the ISM and the mass of stars.
This is the solution of a differential equation (see
\citealt{Edmunds90}) valid in a closed-box case, with instantaneous
recycling, instantaneous mixing, and low metallicities.
The solar yield \ysun, i.e., the fractional contribution of metals to the 
solar mass, is often used as reference and is 0.0126 \citep{Asplund04}.
By inverting this equation, and using observed quantities for $Z$ and \fgas,
we obtain the {\em effective} yields: 
\begin{equation}
y_{eff}= Z/ln(1/f_{gas})
\end{equation}

The measured values of \yeff\ could differ from the true stellar 
yields $y$
if some of the assumptions used to derive 
eq.~\ref{eq:yields}
do not hold, in particular if the system is not a closed box.
Outflows of very enriched material, such as the ejecta from SNe,
extract metals, decreasing $Z$, while infalls of pristine gas can increase
\fgas\ and decrease $Z$. 
It is well known that both effects must be present at some level.
Gas infall, either in merging episodes or in cold gas accretion, is needed to 
supply galaxies with the amount of matter we observed in stars today, while
starbursts are known to produce
SN driven winds of the order of 0.1 \msun yr$^{-1}$ kpc$^{-2}$
\citep{Lehnert96a}. 
The presence of both effects does not imply that both contribute 
significantly to
the chemical evolution of the systems and, in particular, to the
observed yields.
In all cases, \yeff\ must be smaller than
$y$, as discussed, for example, by \cite{Edmunds90} and 
\cite{Dalcanton07}. This is why it is interesting to measure the effective
yields, because it can measure how a system is far from being a closed-box.\\

In the local universe, \cite{Tremonti04} have found a significant 
dependence of \yeff\ on mass, i.e., lower mass galaxies have lower
\yeff. 
The reduction of \yeff\ below the solar value \ysun\ in  galaxies with low stellar mass
is usually described as a consequence of outflows.
If outflows are the main effect shaping the mass-metallicity 
relation, then the effective yields are expected to increase with 
stellar mass because 
lower mass galaxies, having a shallower potential well,
have lost a larger fraction of metals into the 
intergalactic medium. Above a certain stellar mass, the galaxy
potential well is too deep for the SNe to eject significant fraction
of enriched gas, and \yeff$\sim y$ is expected in massive galaxies.

Figure~\ref{fig:yields} shows the values of \yeff\ for the LSD
galaxies.
The errors on each single point are large, as they reflect the large 
uncertainties on both gas fractions and metallicities. Systematic
effects on dimensions could also be present, as discussed above.
Nevertheless, the situation seems pretty clear:
rather than {\em increasing} with stellar mass toward solar values,
in the LSD sample \yeff\ 
is found to {\em decrease} with stellar mass, starting from \ysun\ 
at the low mass end.  
The differences with the results by 
\cite{Tremonti04} are striking. 
This is not surprising because it means that the physical processes
dominating the SDSS galaxies in the local universe
are not the same as in LBGs at $z>$3.
In contrast, our result is very similar to what has been found 
by \cite{Erb06a}. Our and Erb's samples share 
the same dependence of \yeff\ with stellar mass, although LSD galaxies
have lower \yeff\ because of their lower metallicity.
Our results are also similar to those of \cite{Weiner08}
at $z$=1.4 and \cite{Law07} at $z$=2.5, who detected the 
presence of stronger galactic winds
in more massive galaxies.

To interpret this result we have to take into account how the
samples were chosen and the scaling relations.
From the {\em expected} gas fraction in Figure~\ref{fig:gasden}
and the mass-metallicity relation in Figure~\ref{fig:massmet},
we can {\em predict} the \yeff\ of the LSD sample.
This is a real prediction, i.e., once you define SFRs, metallicity
and dimensions,
there are no free parameters to compute yields.
We plot this prediction in Figure~\ref{fig:yields}.
The observed behavior is in agreement with what is expected for 
a SFR-limited sample.
In other words, the dependence
of the yields on stellar mass is a direct consequence of the
mass-metallicity relation and of the other scaling relations.

The presence of yields decreasing with mass implies that the classical 
outflows, whose specific power decreases with increasing galaxy mass,
do not apply to the LBG at $z$$\sim3.1$.
The mass-metallicity relation at $z>$3 must have a different origin. 

%---------------------------------------------------------------------
\begin{figure}
\centerline{\includegraphics[width=\columnwidth]{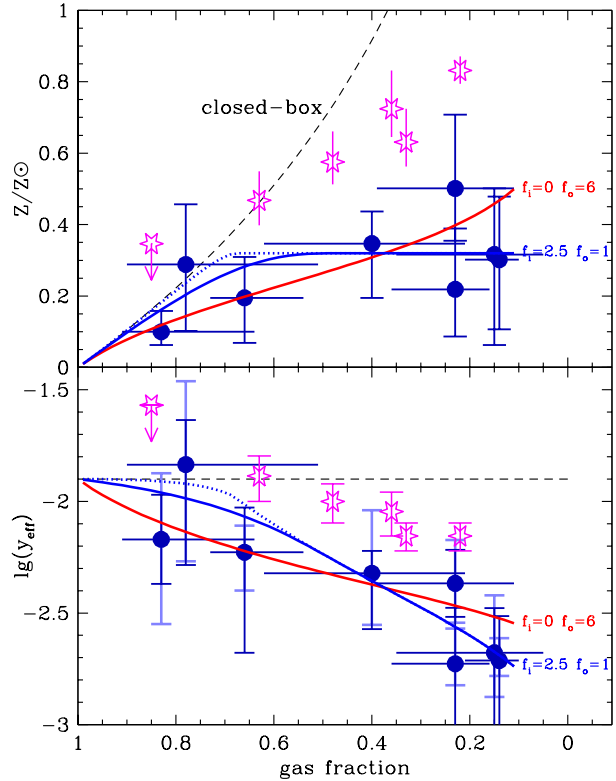} }
\caption{
Metallicity ({\em top}) and effective yields ({\em bottom})
as a function of gas fraction. This is plotted decreasing rightwards, 
to have galaxies evolving from left to right. 
The blue solid dots are the LSD galaxy 
sample, the magenta stars the \citet{Erb06a} data. 
The black dashed line shows the expectations from a closed-box model.
The solid lines show the results of applying the model in \citet{Erb08}
with values of $f_i$ and $f_o$ in the labels. The red line shows a model
with pure outflows, with ejected mass 6 times the SFR. The blue line
is a model with both infall ($f_i$=2.5) and outflow ($f_o$=1). 
The blue dotted line is this same model but without 
outflow ($f_i$=2.5 and $f_o$=0).
}
\label{fig:plotyields}
\end{figure}
%---------------------------------------------------------------------

\smallskip

The dependence of \yeff\ on stellar mass is a secondary effect due to the 
relation between stellar mass and gas fraction. For this reason,
the physical meaning of these findings is better understood by plotting 
\yeff\ directly as a function of gas fraction.  This is done in 
Figure~\ref{fig:plotyields}. The relation between these two quantities
and the effect of infalls and outflows are described by 
several authors. 
In particular, \cite{Dalcanton07} shows that
outflows of very enriched materials can be very effective in reducing \yeff,
especially for gas-rich systems. Gas-poor system can change their \yeff\ 
by a factor of two, as observed in the LSD galaxies, if they accrete a 
significant amount of gas, larger than the internal gas.

For gas rich systems, \yeff\ does not depend
on \fgas\ and it is equal to the stellar yields $y$
\citep{Dalcanton07},
thus allowing us to measure this quantity, which in principle could be different
from \ysun. 
Metallicity can influence nucleosynthesis in different ways. The most important effect
is expected to be mass-loss, which has a strong dependence on metallicity.
Massive stars with stronger stellar winds lose a larger fraction of He and C that
would otherwise be converted to O. As a consequence, the oxygen yield we measure
is expected to decrease with metallicity.
The expected effect is not large, up to a factor of 2, but in principle 
measurable with our data 
\citep{Woosley95,Marigo01,Garnett02}.
All the LSD galaxies with \fgas$>0.7$ have \yeff\ fully consistent with \ysun, showing
that any metallicity dependence of $y$ is not large even at $Z\sim0.1Z_{\odot}$.

\smallskip

\cite{Matteucci01b} (see also \citealt{Matteucci08b}) and
\cite{Erb08} introduce a model in which
galaxies have infalls and outflows proportional to the SFR. 
The instant recycling and mixing approximations are used, 
infalling gas is assumed to have no metallicity, while the metallicity of the
outflowing gas is considered to be the same as the ISM.
This model can be used to reproduce \yeff\ by changing 
two free parameters, i.e., the amount of infall $f_i$ and outflow $f_o$
in unit of the SFR of the galaxy. 
Two more parameters can be varied, 
if necessary, i.e, the true yield $y$, and the fraction $\alpha$ of 
mass still locked in stars. 
Implicitly, the physical properties of these outflows differ
significantly with what is invoked to explain the yields 
in the local universe: given the correlation between SFR and \mstar\ 
\citep{Schiminovich07,Erb06c}, at $z$$\sim$2 these
outflows are expected to increase, rather than decrease, with stellar mass.
\cite{Erb06a} and \cite{Erb08} find that their data at $z$=2.2 can be 
adequately  reproduced by a 
super-solar true yield ($y$=1.5\ysun) and significant infalls 
and outflows, with the 
best-fitting parameters of $f_i$=2.2 and $f_o$=1.3.

We have applied this model to LSD, reproducing the dependence of \yeff\ on \fgas\ 
varying $f_i$ and $f_o$. 
Infalls are effective in changing \yeff\ for the gas-poor galaxies, 
leaving gas-rich galaxies
much more unaffected. Outflows are more efficient in reducing 
\yeff\ in gas-rich galaxies, and much less effective for gas-poor galaxies.
In contrast to \cite{Erb08}, we have fixed
the values of $y$ to \ysun, as explained above. The parameter $\alpha$ is a
slowly decreasing function of time, as more stars leave the main sequence,
but can vary in quite a small range. 
It is expected to be $\sim$0.93 at t=$10^7$yr, $\sim$0.84 at t=$10^8$yr, 
$\sim$0.76 at t=$10^9$yr,
and approaching 0.60--0.65 after an Hubble time \citep{Bruzual03}. 
Given the ages of our galaxies, we have used $\alpha$=0.8, and the results
are not very sensitive to the value chosen. 
The results are shown in Figure~\ref{fig:plotyields}.
Outflows without infalls can explain the observed \yeff, but in this case
large values of $f_o$ between 4 and 8 are needed.
The Figure shows the best-fitting pure outflow model, having $f_o$=6.
In other words, if pure outflows 
are responsible for the reduction of \yeff, the most active
galaxies must eject into the IGM masses of the order of 400--800\msun/yr, while
converting 100\msun/yr into stars. This must happen in galaxies
as massive as $3\cdot10^{10}$\msun, i.e., 
galaxies that, in the local universe, show solar yields.
As a comparison, \cite{Weiner08} found outflowing masses of about
20 \msun/yr in their sample of starburst galaxies at $z$=1.4, whose SFRs
are 10--100 \msun/yr, similar to the present sample.
Pure infalls can explain the yields with lower amount of flowing gas. 
In this case, the best fitting value is $f_i$=2.5, with an acceptable 
range between $f_i$=1.5 and $f_i$=3.5. 
Once an infall with $f_i\sim2$ is present, 
it dominates the behavior of \yeff\ and $f_o$ can assume any value 
between 0 and 7.
The best-fitting model with $f_i$=2.5 and $f_o$=1 is also plotted in 
Fig.~\ref{fig:yields} as a function of stellar mass, providing a good fit.

\smallskip

%==============================================================================
\section{Summary and conclusions: a picture of LBGs }

We have obtained deep, spatially resolved, near-IR spectroscopy of
a complete sample of LBGs at $z$$\sim$3, selected only to be near
a bright foreground star needed to drive the adaptive optics system.
These galaxies are expected to give an unbiased representation
of the \cite{Steidel03} sample of LBGs.
The detected optical lines give information on many aspects of these 
galaxies, and in this paper we have analyzed the properties
related to metallicity, stellar and gas mass, and gas fraction.

A strong evolution of the mass-metallicity relation is found, 
i.e., LBGs at $z$$\sim$3 have metallicities between 0.1 and 0.5 solar.
Assuming that the Schmidt-Kennicutt law is already in place at $z$$\sim3$, 
we derive the properties of the gas. Large masses of gas ($\sim10^{10}$\msun)
are present
in these galaxies, as a result of recent episodes of gas accretion. 
The gas fraction is found to be
anti-correlated with stellar mass, and this is possibly due to
the UV selection of the sample. 
These results on gas mass and metallicity imply that,
at these redshifts, LBGs are already massive but still metal-poor.
This means that mass assembling is occouring before
star formation, as observed in early-type galaxies at lower redshifts
(e.g., \citealt{Saracco03,Saracco05,Daddi05})
Once the selection effects are accurately considered,
this can be used 
to put constraints on the models of galaxy formation and evolution. 

The time scale to produce the observed stars at the current level of 
star formation varies from $\sim10^8$yr for the smallest systems 
to $\sim10^9$yr for the galaxies 
with the largest stellar mass. In contrast, the timescale to exhaust 
the available gas
is less variable, and is comparable with the dynamical times of 
these systems.

The effective yields decrease with
stellar mass and have solar values at the low-mass end. This is in 
contrast with what is observed in the local universe, and can be
ascribed to recent accretion of metal poor gas
at a rate of the order of the current SFR.
Stellar yields have solar values even at Z$\sim$0.1\zsun.

All these evidences show that the LBGs of our sample
can be described as starbursts on several dynamical timescales
following major 
events of gas accretion, due either to gas infall or
merging with gas-rich, metal poor galaxies.
Low mass galaxies 
can be described as experiencing their first major 
burst of star formation. 
More massive galaxies are more gas poor
and have lower yields, as is expected when gas infall ignites
a new episode of star formation in an older galaxy.
Outflows are probably present  but do not dominate
the chemical evolution of these systems, that is more easily
driven by infalls.\\

These galaxies will soon be observed by HST at optical and near-IR bands.
These data will complement our near-IR, spatially resolved spectroscopy
in oder to study the morphological and dynamical properties of these galaxies
and obtain a complete picture of these LBGs.

%=========================================================================
\smallskip
{ \bf Acknowledgments}
We thanks F. Eisenhauer and A. Modigliani for support during data reduction,
and B. Kennicutt and T. Nagao for very useful discussions.
We also thanks the staffs of ESO, Spitzer and TNG for excellent 
service observing, and E. Daddi for having provided data in tabular form.
This work was partially supported by the Italian Space Agency
through contract ASI-INAF I/016/07/0, by INAF CRAM 1.06.09.10,
and by NASA (Spitzer) grant number 1343503.

%-----------------------------------------------------------------------------

\bibliographystyle{/Users/filippo/arcetri/Papers/aa-package/bibtex/aa}
\bibliography{/Users/filippo/arcetri/bibdesk/Bibliography}

\end{document}